# From an Integrated Usability Framework to Lessons on Usability and Performance of Open Government Data Portals: A Comparative Study of European Union and Gulf Cooperation Council Countries

Fillip Molodtsov University of Tartu, Estonia fillip.molodtsov@ut.ee
Anastasija Nikiforova University of Tartu, Estonia, nikiforova.anastasija@gmail.com

**Abstract:** Open Government Data (OGD) initiatives aim to enhance public participation and collaboration by making government data accessible to diverse stakeholders, fostering social, environmental, and economic benefits through public value generation. However, challenges such as declining popularity, lack of OGD portal usability, and private interests overshadowing public accessibility persist. This study proposes an integrated usability framework for evaluating OGD portals, focusing on inclusivity, user collaboration, and data exploration. Employing Design Science Research (DSR), the framework is developed and applied to 33 OGD portals from the European Union (EU) and Gulf Cooperation Council (GCC) countries. The quantitative analysis is complemented by qualitative analysis and clustering, enabling assessment of portal performance, identification of best practices, and common weaknesses. This results in 19 high-level recommendations for improving the open data ecosystem. Key findings highlight the competitive nature of EU portals and the innovative features of GCC portals, emphasizing the need for multilingual support, better communication mechanisms, and improved dataset usability. The study stresses trends towards exposing data quality indicators and incorporating advanced functionalities such as AI systems. This framework serves as a baseline for OGD portal requirements elicitation, offering practical implications for developing sustainable, collaborative, and robust OGD portals, ultimately contributing to a more transparent and equitable world.

**Keywords:** Design Science Research, Open government data, data portal, Open data ecosystem, Usability, European Union, Gulf Cooperation Council

## 1    Introduction

A fundamental principle of open government is the transparency of governmental information and operations, which underpins the pillars of *participation* and *collaboration* (Obama, 2009). Open Data and Open Government Data (OGD) exemplify this transparency, offering platforms for public engagement and collaboration across diverse array of stakeholders, empowering users to create applications, services, and insights with social, environmental, and economic benefits, driving innovation and collective problem-solving (Susha et al., 2015; Reggi & Dawes, 2022; Ruijer, 2021; Purwanto et al.,2020a; Papageorgiou & Charalabidis, 2023; Fang et al., 2024). However, despite its initial success, the OGD movement faces challenges, including declining popularity and concerns about data accessibility overshadowed by private interests commonly referred to as "data winter," highlighting the urgent need to establish a data ecosystem that views not as a commodity to be traded but as a resource that empowers communities and advances scientific research, contributing to a more informed and equitable world (Verhulst, 2024).

Open data portals serve as pivotal gateways to OGD playing a crucial role in this ecosystem and aiming to foster civic engagement and business opportunities while ensuring accessibility for users of all backgrounds (Janssen, et al., 2012; Shen, & Vlahu-Gjorgievska, 2024). Research indicates that OGD portals have the potential to foster business opportunities, particularly for Small and Medium-sized Enterprises (SMEs), and to enhance civic engagement (Zhu & Freeman, 2019). Yet, challenges persist, including portal usability, communication with diverse populations, and strategic value creation (Carsagina et al., 2022; Nikiforova, 2020b; Dawes et al., 2016; Schwoerer, 2022; Benmohamed et al., 2024; Aarshi et al., 2018; Wang et al., 2018; Lněnička et al., 2021; Nikiforova & Lnenicka, 2021; Ansari, Barati, & Martin, 2022; Benmohamed et al., 2024).

The European Commission's Open Data Maturity Report (ODM report) (Page et al., 2023) found that 11 countries (almost half of EU members) scored 90% or above on the portal dimension, with Poland, Estonia, and Ireland receiving the highest scores. However, a high ranking does not imply perfection - it is only relative quality. The ODM reports, however, are based on self-reports provided by representatives of the OGD initiative (Carsagina et al., 2022), which may raise concerns regarding the credibility of the results and suggests that the user's perspective may be omitted from its assessment as it is not the purpose of this index (Lnenicka, Luterek, & Nikiforova, 2022). Excluding users without domain knowledge from the data ecosystem has been a significant and common concern in recent years (Nosheen et al., 2019).

To accelerate the development of user-friendly, collaborative, robust, and sustainable portals, it is crucial to identify the current state of the art and best practices that these portals should adhere to. This involves considering current trends not only in the field of open data but also in software engineering (SE), human–computer interaction (HCI), and user experience (UX). Due to the dynamic nature of continuous development, many frameworks quickly become outdated or limited. This obsolescence hampers the evaluation of portals and the creation of a sustainable agenda for their improvement, preventing the implementation of a portal that meets user needs and expectations. Numerous indexes and benchmarks have been proposed in the literature to evaluate OGD efforts, including OGD portals (Máchová et al., 2018; Carsaniga et al. 2022; Sieber & Johnson, 2015; Afful-Dadzie & Afful-Dadzie, 2017; Matheus et al., 2021; Zuiderwijk et al., 2014). However, research by (Kao, 2023) suggests that future studies should integrate these different frameworks. While specific target areas of OGD benchmarks can be assessed separately, an integrated framework would be beneficial. Additionally, benchmarks encompassing a large number of geographic regions are necessary. This study addresses both of these gaps by responding to the call for such integrated and comprehensive research.

The objective of this study is to propose an integrated framework for evaluating the usability of OGD portals. Based on the research, the proposed framework is centered around three dimensions that were found by the research to be key for a inclusive, resilient and sustainable OGD portals: (1) inclusivity, ensuring the portal is accessible to a wide range of users, including both local/internal and external users of different nationalities and countries being available in different languages; (2) supporting and facilitating user collaboration and active involvement/participation; and (3) facilitating exploration and understanding of data.

Methodologically, Design Science Research (DSR) is employed, encompassing information collection, framework prototyping, portal assessment, framework compilation, and result analysis stages. Through cluster analysis, relationships and patterns among portals are discerned based on performance metrics.

By refining criteria and metrics for OGD portal evaluation, incorporating recent literature trends, and shedding light on underexplored GCC and EU OGD portals, this study offers practical implications for stakeholders to develop sustainable, collaborative, and robust portals. Recommendations derived from the analysis contribute to the ongoing discourse on open data portal efficacy and quality benefiting a wide range of stakeholders in the open data ecosystem. This paper extends our study [anonymized] presented at [anonymyzed conference], by delving deeper into the framework's development through the DSR cycles, its application to selected portals. It presents dimension-wise results with identified best practices and common weaknesses, and provides recommendations and implications for future research.

The rest of the paper is structured as follows: Section 2 provides the research methodology, Section 3 provides a brief overview on the state of the art, Section 4 presents the developed framework, Section 5 presents the results of applying the framework to selected portals along with the cluster analyses based on the score matrix, Section 6 provides a discussion of the results, recommendations, as well as the limitations of the study, and Section 7 concludes the paper.

## 2    Methodology

To achieve the objective of this study, which is the development of an integrated framework for evaluating the usability of open data portals, the Design Science Research (DSR) methodology (Hevner, 2007) is employed. Grounded in rigorous scientific inquiry, DSR is a research paradigm used in information systems and computer science to address complex problems by designing and evaluating innovative artifacts, combining problem-solving with the creation of novel solutions, and emphasizing iterative refinement through Relevance, Design, and Rigor cycles (see 1)(Hevner, 2007).

Following this methodology, the integrated framework is developed during the Design Cycle, aligning with the knowledge base in the Rigor Cycle and the environment in the Relevance Cycle, and then tested against portals. The research progresses through five main stages: information and requirements collection (Stage 1, Rigor Cycle), framework prototyping (Stage 2, Design Cycle), portal assessment to test the prototype (Stage 3, Relevance Cycle), framework compilation (Stage 4, Design Cycle), and framework testing and result analysis (Stage 5, Relevance Cycle).

In Stage 1, a systematic literature review (SLR) is conducted. Stage 2 involves compiling a draft version of the framework and conducting preliminary review. Stage 3 entails testing the developed artifact - framework - against a small subset of OGD portals to identify potential improvements. Stage 4 involves considering suggestions for enhancement, refining the framework, and conducting a final review. In Stage 5, the integrated framework is tested on 33 national OGD portals, including 27 EU and 6 GCC portals, with qualitative and quantitative analysis, including cluster analysis based on generated quantitative data. Recommendations are then formulated based on the results and analyses.

Central to DSR is the collaboration between researchers and practitioners, ensuring that the developed artifacts contribute both to theoretical knowledge and practical utility in real-world contexts. In this study, direct collaboration with practitioners was omitted. However, the resulting integrated framework was repeatedly reviewed by two experts in the field of open data (OD) portals and user experience (UX). Within this study, we define an individual to be expert if individual has expertise in computer science and information systems, possessing over 5 years of experience in software engineering (SE), Human-Computer Interaction (HCI), or User Experience (UX) projects, and extensive expertise in Open Government Data (OGD) research.

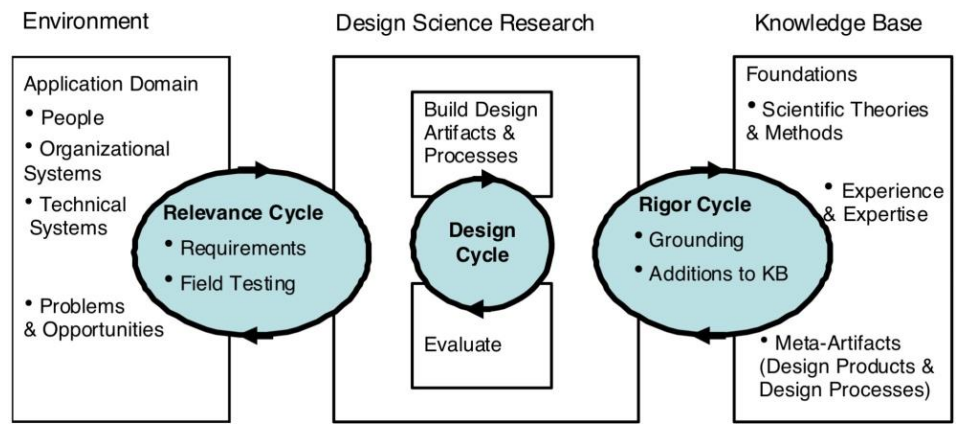

Figure 1. Design Science Research Methodology

## 2.1  Information and requirements collection

In Stage 1, a systematic literature review (SLR) is conducted, combining findings from the SLR with feedback from leading experts on OGD and OGD portal design principles in particular (Janssen et al., 2012; Purwanto,

2020a,b,c; Matheus et al., 2021).

The raw data obtained through the SLR (that resulted in 82 studies) was systematically analyzed, extracting portal features related to user usability and compliant with the framework's scope to build the framework upon. As SLR was presented in (Molodtsov and Nikiforova, 2024) with supplementary data (list of articles, protocol etc.) made available on Zenodo - https://doi.org/10.5281/zenodo.10985804, we do not repeat its description here. In addition, a list of commonly overlooked usability features has been collected and categorized, along with more non-trivial suggestions for improving OGD portal usability beyond refining missed features.

## 2.2  Prototyping the framework

The framework design process involves several steps. First, the data collected for each metadata dimension from the selected articles (available on Zenodo) is analyzed to derive "patterns" for both sub-dimensions and dimensions within the framework. These patterns are then reviewed and prioritized, forming the basis for prototyping the draft framework. Alongside dimensions and sub-dimensions, criteria for evaluating the adequacy of sub-dimensions are also specified.

The prototype framework, including its aspects (sub-dimensions) and dimensions, was reviewed by 2 experts. The first expert is a master's student who works in the private sector, with over 4 years of experience in SE projects and OGD expertise in research. The second is a PhD holder with over 5 years in academia and practice in SE, HCI, and UX projects, and OGD expertise in research and public administration consultations. This review assessed the sub-dimensions consistency and conformance to the heuristic evaluation method (Nielsen & Molich, 1990).

## 2.3  Conducting portal assessment to test the prototype

During Stage 3, the framework prototype undergoes testing to determine if additional sub-dimensions should be incorporated or if existing sub-dimensions can be further refined. This testing is conducted on four top-performing national Open Government Data (OGD) portals, namely French, Irish, Estonian, and Spanish, as identified by the ODM report 2022 (Carsaniga et al., 2022), which was the most recent edition at the time of framework design. These portals were selected based on the assumption that they implement numerous best practices and showcase emerging trends.

The testing process was exploratory, involving an examination of the framework's structure and rationale within the context of live portals. Through comparing the selected portals, it became evident that adjustments such as adding new sub-dimensions, refining sub-dimension criteria, or subdividing existing sub-dimensions were necessary. Each time a portal underwent testing, any modifications made as a result of its evaluation were documented, and the affected sub-dimensions of previously evaluated portals were reassessed. The prototype testing concluded once no further modifications were required, and the performances of the portals were re-evaluated accordingly.

## 2.4  Compiling the framework

In Stage 4, all available information sources (SLR, selected articles from experts in portal design, notes from framework prototype testing and all suggestions and insights gathered (incl. from the reviewers of the conference article (Molodtsov & Nikiforova, 2024) were thoroughly reviewed, leading to the development of the final version of the framework. Similar to Stage 2, the final version of the framework underwent a rigorous review to ensure its completeness and accuracy.

During this stage, it became evident that the dimensions and sub-dimensions varied significantly in terms of their nature and importance, necessitating the implementation of a weighing system. Subsequently, after finalizing the framework, a weighing system for score calculation was devised. This weighing system was developed by considering existing systems encountered in selected publications from experts in portal assessment framework design (Susha et al., 2015; Zuiderwijk et al., 2021; Máchová et al., 2018; Alexopoulos et al., 2018; Afful-Dadzie & Afful-Dadzie, 2017; Carsaniga et al., 2022).

## 2.5   Framework testing and result analysis

In Stage 5, the developed framework underwent testing using a sample comprising 33 national Open Government Data (OGD) portals, encompassing 27 EU portals and 6 GCC portals.

The inclusion of EU and GCC national Open Government Data (OGD) portals in this study highlights the competitive landscape and legislative initiatives driving EU portals, while addressing the gap in understanding the current status of GCC portals. By applying the proposed framework to these selected portals, we aim to rank OGD portals, identifying the most competitive ones, highlight best practices that lower-performing portals can adopt, and pinpoint common shortcomings.

Conversely, GCC national portals have not been sufficiently studied. Although some research has covered GCC portals using existing frameworks (Máchová et al., 2018), these analyses are outdated due to rapid technological advancements, leaving the current status of GCC OGD portals unclear. Cooperation between the EU and GCC countries is crucial for achieving political and economic goals (Meissner, 2019). Comparing OGD portals from these regions can determine their comparability and potential for cross-border (and interregional) interoperability, thereby (indirectly) strengthening political and economic ties.

A systematic web search methodology was employed to identify relevant portals for each country. This involved utilizing the country name in combination with English search terms such as "*open data*" and "*open data portal*" on the Google search engine. In cases where the official status of a particular portal was uncertain or inconclusive, an inventory of open data portals (Herrmann & Robbins, 2023) was consulted. Alternatively, the web addresses of EU national OGD portals were obtained by referencing the ODM report. Web addresses of the portals used in the testing process are provided in Appendix I.

As a result of portal testing:
- each individual portal undergoes assessment, with scores for sub-dimensions, dimensions, and total score calculated using the weighing system;
- average scores are computed for EU and GCC portals;
- portals are ranked based on their performance;
- best-performing portals for each dimension are identified, elucidating observed best practices and weak points on national portals, contributing to qualitative analysis alongside quantitative analysis;
- trends in portal design are discerned, and collaborative initiatives between portals are identified.

To gain deeper insights into the relationships and patterns among different portals based on their performance metrics, clustering analysis is employed. This technique, fundamental in data mining and machine learning, organizes the dataset into groups (clusters) based on similarities between data points, facilitating the identification of patterns within the data (Govender & Sivakumar, 2020). Two types of clustering analysis, K-means clustering and hierarchical clustering, are conducted based on the score matrix. K-means clustering partitions data into a predefined number of clusters by minimizing within-cluster variance, while hierarchical clustering constructs a tree-like hierarchy of clusters by iteratively merging clusters based on their similarity (Govender & Sivakumar, 2020).

For K-means clustering, the optimal number of clusters is determined using the Elbow method, with the decision guided by optimal boundary differentiation and maintaining a reasonably small number of clusters. In hierarchical clustering, the Ward method is utilized to build the linkage matrix, and the distance threshold for cluster formation is selected by analyzing the dendrogram of the linkage matrix. The distance threshold is chosen to ensure equivalence in the number of clusters generated by both clustering methods.

Evaluation of the average dimensional scores of portals from both types of clusters provides insights into their performance across multiple dimensions. The corresponding cluster of the other type is selected based on the greatest number of common portals, facilitating comparative analysis.

# 3 Results of the literature review

As a result of the SLR (Molodtsov & Nikiforova, 2024), we found that in 2021, there was a notable surge in published articles, many of which emphasized the crucial importance of enhancing data accessibility to address pressing global challenges such as COVID-19 (Krismawati & Hidayanto, 2021; Harrison et al., 2021; Nikiforova, 2021). This trend supports the assumption that there is a need to (1) develop an integrated framework aligned with recent developments, (2) reassess portals to instigate necessary changes based on identified shortcomings, and (3) undertake a comprehensive evaluation spanning across borders and regions (Commision, n.d., 2023).

Most studies (47) have examined portals within the European Union, with Asia following closely with 25 studies, and North America with 24 studies South America received less attention with 13 studies followed closely by Africa and the Pacific with 14 studies each, It is noteworthy that studies typically encompass portals from various regions rather than focusing solely on one area. Research indicates that the most robust OGD portals are typically associated with countries from Europe, North America, East Asia, and Australasia.

The literature offers a spectrum of frameworks, ranging from conceptual and high-level to detailed frameworks (Kubler et al., 2018; Máchová & Lnénicka, 2017; Lnenicka & Nikiforova, 2021). The Open Data Maturity (ODM) reports by the European Commission, demonstrate how frameworks evolve to incorporate new trends, adding and removing criteria over time (Carsaniga et al., 2022; Knippenberg, 2020). While conceptually similar, these frameworks are challenging to compare directly (Susha, et al., 2015; Zuiderwijk & Susha, 2021). Consequently, researchers often resort to reusing established frameworks that suit their specific needs (Mutambik et al., 2021). The usability framework by Machova et al. (2017) is particularly prevalent, despite its age and limited coverage of contemporary trends. In contrast, the transparency-by-design framework (Lnenicka & Nikiforova, 2021) offers broader coverage but still lacks in usability aspects, necessitating enhancements to address current trends like gamification and sustainability (Simonofski et al., 2022; Lnenicka et al., 2024a). Among the studies analyzed, several frameworks have been reused, including those focused on usability and transparency, although none fully capture the intricacies of modern OGD portals. These insights inform the development of an integrated framework tailored to assess the state of usability in contemporary OGD portals.

The literature highlights several recurring deficiencies and recommendations for improvement across open data portals. Commonly cited deficiencies in general portal features include *poor navigation, information overload, lack of prioritization in displaying information, multilingualism, accessibility features, metadata absence or low metadata quality, inconsistency, data versioning issues, inadequate search functionality, unprocessed data download limitations, scarcity or low value of datasets, absence of data visualization and analytical tools*, and *inadequate feedback and support mechanisms* (Mutambik et al., 2021; Sisto et al., 2018; Gill & Corbett, 2017; Ferati et al., 2020; Weerakkody et al., 2017; Nikiforova & McBride, 2021; Alexopoulos et al., 2018; Elsawy & Shehata, 2023; Sheaoran et al., 2023; Nikiforova & Lnenicka, 2021; Dahbi et al, 2018; Wang & Shepherd, 2020; Herrera-Melo & González-Sanabria, 2020; Knippenberg, 2020). Concerns about portal sustainability include the absence of a strategy and performance dashboards (Reggi, 2020; Lnenicka et al., 2024a; Quarati & De Martino, 2019).

Recommendations for addressing these issues include *implementing missing features* or fixing found issues, with some more sophisticated, e.g., *limiting free-form fields for metadata, providing predefined options, keeping dataset descriptions concise, minimizing registered-user-only actions, using the DCAT-AP vocabulary for metadata standardization and enhanced and interoperability, migrating to advanced technological platforms like CKAN, DKAN, and Socrata, understanding user needs and demands, attracting a wider audience, reducing information pollution on dataset pages, implementing query recommendation systems and automatic dataset description filling, supporting multilingualism, focusing on lay citizens, introducing gamification elements,* and *using storytelling* to enhance content understandability (Nikiforova, 2020a; Máchová et al., 2018; Klimek, 2019; Zhu & Freeman, 2019; Simonofski et al., 2022; Lnenicka et al., 2024a; Máchová & Lnénicka, 2017; Zhu & Freeman, 2019; Nikiforova & McBride, 2021; Schauppenlehner & Muhar, 2018; Kacprzak, Koesten, Ibáñez, Simperl, & Tennison, 2017; Kacprzak et al., 2019; Nikiforova & Lnenicka, 2021).

The identified missing features and recommendations are integral to the development of the framework under consideration.

# 4    Proposed integrated usability framework for evaluating OGD portal usability

The proposed integrated usability framework focuses on: (1) inclusivity, ensuring the portal is accessible to a wide range of users, including both local/internal and external users of different nationalities and countries being available in different languages; (2) supporting and facilitating user collaboration and active involvement/ participation; and (3) facilitating exploration and understanding of data.

The proposed integrated usability framework emphasizes (1) *inclusivity*, ensuring accessibility for a diverse range of users, including local and international users, with multilingual support, (2) *user collaboration and participation*, facilitating active user involvement, (3) *data exploration and understanding*, enhancing users' ability to explore and comprehend data.

The framework (see Tables 1-2, and Zenodo) comprises 9 dimensions divided into 72 sub- dimensions:

- multilingualism (4 sub-dimensions), which refers to interface availability, content availability, and search functionality in different languages;

- navigation (3 sub-dimensions), which focuses on user interface elements such as menu structures, bread-crumbs, and tabs to facilitate navigation;

- general Performance (4 sub-dimensions), which covers load time, responsive design, error-free experience, and overall accessibility;

- data understandability (11 sub-dimensions), which encompasses high-value datasets (HVD), dataset views and downloads, dataset re-use, data previews, visualization tools, and simplified content for better comprehension;

- data quality (9 sub-dimensions), which includes machine-readable formats, metadata elements, update frequency accuracy, temporal and spatial coverage, quality ratings and associated explanations, and an automated quality check;

- data findability (15 sub-dimensions), which covers data discoverability by publisher, categories, formats, tags, licenses, sorting options, metadata, API and SPARQL endpoints, recommender systems, and featured topics;

- public engagement (13 sub-dimensions), which refers to availability of use-case uploads feature, community-sourced content, social media integration, notification systems, event promotion, personalization options, request forms, tracking, and gamification elements such as badges, rewards, quizzes, and competitions;

- feedback mechanisms and service quality (7 sub-dimensions), which includes portal-wide comment sections, forums, direct publisher-user communication, dataset-specific feedback, usefulness assessment, guidelines, tutorials, support contact options, and suggestion forms;

- portal sustainability and collaboration (6 sub-dimensions) that refers to sustainability strategies, performance dashboards, collaboration with regional and international governments, user satisfaction surveys, and use of open-source code.

To assess the presence of specific aspects, a Boolean evaluation (1/0) is primarily used, with additional notes taken (if any) for qualitative analysis. For accessibility (c4), the web-based accessibility checker (AccessibilityChecker.org, 2023) is used; a portal scores 1 if it achieves 71% or higher, indicating compliance without critical issues.

Sixteen sub-dimensions/aspects, namely d2-5,7-8,10-11, e1-3,7, f10-11 are evaluated on a sample basis with a threshold of 70%, i.e. 10 out of 14 datasets, to achieve 1 point. The grading of e4 - dataset update frequency accuracy - is tied to e3 - dataset update frequency, checking if at least 70% of datasets' update frequencies are accurate. For example, if a dataset's update frequency is "monthly," the latest modification date should be the current or previous month. If the frequency is specified but unverifiable, it is not considered fulfilled.

For sub-dimension assessment based on a dataset sample, the sample is created as follows: if the portal supports sorting by relevance (popularity) and modification date, the first four and last three datasets from the data catalog list form the sample. If only sorting by modification date is available, the first eight and last six datasets are used. If no sorting is implemented, the first eight and last six datasets are taken.

Despite efforts to classify aspects according to their primary dimension, some elements may belong to different dimensions or serve significant roles in other dimensions. Aspects also vary in importance to the framework's core ideas, necessitating a weighting system. To this end, we consulted literature to identify potential weighting systems (see Table 3).

Popular approaches include using (1) equal weights for dimensions and aspects (Lnenicka & Nikiforova, 2021; Zhu & Freeman, 2019; Máchová & Lnénicka, 2017), (2) equal weights for dimensions but different for sub-dimensions (Sisto et al., 2018; Raca et al., 2021; Susha, Zuiderwijk, et al., 2015), (3) different weights for dimensions and aspects (Carsaniga et al., 2022; Knippenberg, 2020; Herrera-Melo & González-Sanabria, 2020; D. Wang, Chen, & Richards, 2018), (4) different weights for dimensions (Afful-Dadzie & Afful-Dadzie, 2017; Zuiderwijk et al., 2021), (5) importance- /priority-based (Susha et al., 2015). While the use of equal weights for dimensions and aspects is one of the most popular approaches due to its simplicity, this approach is often criticized even by those who use it (e.g., (Lnenicka et al., 2022)). Therefore, instead, we use a a priority-based option (similar to (Susha et al., 2015)), where each aspect's score is multiplied by its importance to the framework's central concepts. Three levels of importance—low, medium, and high—are mapped to values of 1, 2, and 3, respectively.

# Table 1. Proposed integrated OGD portal usability framework (1/2)

| Dimension | Sub-dimension | Description | Criteria | Weight |
|---|---|---|---|---|
| (a) Multilingualism | (1) English is one of the supported languages | one of the supported languages is English | 0 - no, 1 - yes | 1 |
| | (2) portal interface is available in the supported languages | content is at least partially translated, ensuring accessibility and usability for a diverse range of users | 0 - no, 1 - yes | 2 |
| | (3) portal content is available in the supported languages | blogposts, dataset pages, manuals, tutorials are translated to the portal-supported languages, enhancing accessibility | 0 - no, 1 - yes | 3 |
| | (4) dataset search can be done in English | users can search for datasets in English (the language of international communication), enhancing accessibility and search functionality | 0 - no, 1 - yes | 3 |
| (b) Navigation | (1) convenient menubar structure | users can easily navigate the website | 0 - no, 1 - yes | 3 |
| | (2) breadcrumb usage | users are provided with a navigational trail, displaying the hierarchical path and aiding in understanding a webpage's location within a website | 0 - no, 1 - yes | 2 |
| | (3) tabs for content-rich pages | content-rich webpages are structured with tabbed navigation elements to facilitate organized presentation and easy access to various sections of information | 0 - no, 1 - yes | 2 |
| (c) General performance | (1) portal loads in less than 4 seconds | the portal should load in 4 seconds | 0 - no, 1 - yes | 3 |
| | (2) responsive web design | functions effectively and is accessible on smartphones and tablets, ensuring a user-friendly experience for mobile users | 0 - no, 1 - yes | 1 |
| | (3) no blocking errors or exceptions | within the basic usage of the portal, no unexpected errors or exceptions were encountered | 0 - no, 1 - yes | 2 |
| | (4) sufficient accessibility level | the accessibility testing is assessed by accessibility checker service (AccessibilityChecker.org, 2023). European Accessibility Act (EAA) compliance was chosen to test against. The threshold to be compliant is a score of 85 | 0 - the score is under 61 (less than 71% of partial compliance without critical issues), 1 - the score is above or equal to 61 | 2 |
| (d) Data understandability | (1) HVD promotion | the promotion of HVD (High-Value Datasets) in portals involves highlighting datasets that are particularly valuable, relevant, or significant for users and encouraging their access and utilization | 0 - no, 1 — yes | 3 |
| | (2) dataset views | the number of times a dataset has been accessed and viewed by users on a data portal or platform | 0 - no, 1 - yes (10/14) | 2 |
| | (3) dataset downloads | the number of times a dataset has been downloaded by users on a data portal or platform | 0 - no, 1 - yes (10/14) | 2 |
| | (4) dataset re-use/showcase count | the number of instances where a dataset has been utilized or repurposed by users on a data portal or platform for various applications or analyses | 0 - no, 1 - yes (10/14) | 1 |
| | (5) re-use/showcase display in dataset page | presenting or showcasing instances where data from a dataset has been utilized or repurposed on the dataset's page | 0 - no, 1 - yes (10/14) | 2 |
| | (6) re-use page dataset list | re-uses are supplied with the list of used datasets | 0 - no, 1 - yes | 3 |
| | (7) data preview | preview of a dataset, offering a sample or snapshot of its content to provide users with a quick understanding of its structure, format, and potential value. | 0 - no, 1 - yes (10/14) | 3 |
| | (8) data visualization, analytics, and filtering tools | software features enable users to visually represent data, analyze it for insights, and refine the information displayed by applying various filters, enhancing data exploration and decision-making | 0 - no, 1 - yes (10/14) | 3 |
| | (9) interactive data visualization | dynamic and user-engaging graphical representations of data enable users to visually represent data, analyze it for insights, and refine the information displayed by applying various filters, enhancing data exploration and decision-making | 0 - no, 1 — yes | 3 |
| | (10) data visualization download | the portal enables saving the results of data visualization or data preview | 0 - no, 1 - yes (10/14) | 2 |
| | (11) vulgarized content(described through examples and visual aid) | simplified or easily understandable description of complex content is provided, making it accessible and relatable to a broader audience without diminishing its quality or value | 0 - no, 1 - yes (10/14) | 3 |
| (e) Data quality | (1) machine-readable data formats | each dataset is available in machine-readable formats | 0 - no, 1 - yes (10/14) | 3 |
| | (2) basic metadata elements | dataset is supplied with metadata consisting of at least: title, description, category, publisher, license, modification date | 0 - no, 1 - yes (10/14) | 3 |
| | (3) update frequency of datasets | the update frequency is specified for datasets | 0 - no, 1 - yes (10/14) | 3 |
| | (4) dataset update frequency accuracy (actual vs promised) | the dataset update frequency accurately reflects the scheduled frequency at which data within the dataset is refreshed or modified, ensuring users are informed about the dataset's currentness | 0 - no, 1 - yes (70%+ of datasets with update fre- quency) | 3 |
| | (5) data temporal coverage | datasets, when appropriate, have the range of time included in them, that is marked in the dataset page | 0 - no, 1 - yes | 2 |
| | (6) data spatial coverage | datasets, when appropriate, have the geographic area for which this dataset is relevant, that is marked in the dataset page | 0 - no, 1 — yes | 2 |
| | (7) dataset quality rating | data quality rating of each dataset is provided | 0 - no, 1 - yes (10/14) | 3 |
| | (8) rating explanation | the criteria (metadata fullness, availability of certain metadata points) for the rating can be found, which provides a deeper insight on a dataset | 0 - no, 1 — yes | 3 |
| | (9) automated dataset quality checklist | an automated dataset quality checklist brings substantial benefits to portal users by swiftly and accurately assessing dataset quality, ensuring that the data they access is reliable, up-to-date, and of the highest standard, enhancing their overall experience | 0 - no, 1 — yes | 3 |
| (f) Data findability | (1) discoverability by publisher | datasets can be filtered by publishers | 0 - no, 1 — yes | 3 |
| | (2) discoverability by categories | datasets can be filtered by categories | 0 - no, 1 — yes | 3 |
| | (3) discoverability by formats | datasets can be filtered by formats | 0 - no, 1 — yes | 3 |
| | (4) dataset format list | list of formats available for the dataset are visible from the catalog (e.g. in the form of tags), providing quick information about the data retrieval methods | 0 - no, 1 — yes | 2 |
| | (5) discoverability by tags | datasets can be filtered by tags | 0 - no, 1 — yes | 2 |
| | (6) discoverability by license | datasets can be filtered by license | 0 - no, 1 — yes | 1 |
| | (7) sorting by modification date | dataset sorting that enables users to organize and arrange datasets based on modification date, facilitating efficient data discovery and access | 0 - no, 1 — yes | 3 |
| | (8) sorting by relevance | dataset sorting that enables users to organize and arrange datasets based on relevance, facilitating efficient data discovery and access | 0 - no, 1 — yes | 3 |
| | (9) sorting by dataset metadata | sorting that enables users to organize and arrange datasets based on metadata criteria, facilitating efficient data discovery and access | 0 - no, 1 — yes | 2 |
| | (10) dataset tags | datasets have descriptive labels or keywords added to enhance searchability and categorization | 0 - no, 1 - yes (10/14) | 2 |
| | (11) data download | data is directly accessible for download on the portal, eliminating the need for users to navigate to external sources or websites for data retrieval | 0 - no, 1 - yes (10/14) | 3 |
| | (12) API endpoints | API endpoints are available | 0 - no, 1 — yes | 3 |
| | (13) SPARQL endpoints / RDF files | SPARQL endpoint or linked data is available | 0 - no, 1 — yes | 3 |
| | (14) recommender system | users are provided with preferably personalized suggestions or recommendations for datasets based on their preferences, interests, and prior interactions with data, making it easier for them to discover relevant and valuable information | 0 - no, 1 — yes | 3 |
| | (15) featured topics | users are offered pages that compile curated data collections related to a specific theme or concept to show a contextual bigger picture | 0 - no, 1 — yes | 3 |

# Table 2. Proposed integrated OGD portal usability framework (2/2)

| Dimension | Sub-dimension | Description | Criteria | We |
|---|---|---|---|---|
| (g) Public engagement | (1) use-case upload feature | users are provided with the opportunity to submit use-cases | 0 - no, 1 - yes | 3 |
| | (2) community-sourced / citizen-generated data | users are allowed to upload community-sourced or citizen-generated data to the portal | 0 - no, 1 - yes | 2 |
| | (3) social media support | links to its official pages on popular social media platforms such as Facebook, X (formerly Twitter), LinkedIn, or others are provided. These links facilitate communication, updates, and engagement between the open data portals and their user communities through social media channels | 0 - no, 1 - yes | 3 |
| | (4) notification system | enables users to subscribe to receive notifications or newsletters, keeping them informed about updates, news, and relevant content pertaining to the portal's activities and offerings | 0 - no, 1 - yes | 2 |
| | (5) portal up-to-date information | the information on the portal is current and up-to-date, ensuring that users have access to the most recent and relevant content | 0 - no, 1 - yes | 3 |
| | (6) sessions and events promotion | the portal provides information about meetings, workshops, or gatherings designed to raise awareness, provide training, or engage the public in using and benefiting from the national portal | 0 - no, 1 - yes | 3 |
| | (7) personalization features | additional features to non-publishers, offering enhanced functionalities and capabilities, are provided. Examples of those features could be: a personalized list of favorite datasets, subscription to topics, comment mentioning system with e-mail notifications, badge collecting | 0 - no, 1 - yes | 1 |
| | (8) badges | virtual achievements or symbols awarded to users for completing specific tasks or reaching milestones, adding an element of accomplishment and recognition to the portal are provided | 0 - no, 1 - yes | 3 |
| | (9) rewards | users are provided with tangible or virtual incentives to encourage desired behaviors and participation, community support | 0 - no, 1 - yes | 1 |
| | (10) quizzes | interactive assessments are provided, engaging users in knowledge testing, promoting learning and user engagement through questions and challenges | 0 - no, 1 - yes | 3 |
| | (11) competition | users compete against each other to achieve specific goals or rankings, fostering engagement and motivation | 0 - no, 1 - yes | 2 |
| | (12) request forms | web-based tools that users can utilize to formally request specific datasets from the data providers or the data portal itself are provided | 0 - no, 1 - yes | 3 |
| | (13) request tracking | page that allows users to monitor the progress and status of their requests, providing them with real-time updates and information about the handling of their inquiries or demands | 0 - no, 1 - yes | 3 |
| (h) Feedback mechanisms and service quality | (1) portal-wide comment sections or forums | online discussion areas where users can engage in conversations, share information, or express their opinions on various topics or aspects of the portal | 0 - no, 1 - yes | 3 |
| | (2) direct publisher-user communication | users can engage in direct and immediate interaction with publishers, facilitating feedback, questions, or discussions. | 0 - no, 1 - yes | 3 |
| | (3) comment sections or forums for datasets | online spaces associated with individual datasets, where users can post comments, questions, and discussions related to that specific dataset, fostering communication and collaboration around the data | 0 - no, 1 - yes | 3 |
| | (4) dataset usefulness assessment | feature allowing users to mark a dataset as useful, often through actions like upvoting or liking, enhances user engagement and provides a way to highlight valuable datasets | 0 - no, 1 - yes | 3 |
| | (5) guidelines, tutorials, manuals, FAQs | informative resources designed to help users understand and utilize a service or product effectively, providing guidance, instructions, and answers to common questions. | 0 - no, 1 - yes | 3 |
| | (6) contact for support | means for users to get in touch with a support team or customer service to seek assistance, ask questions, or report issues related to a product or service | 0 - no, 1 - yes | 3 |
| | (7) improvement suggestion form | suggestion for improvement form is a tool that allows users to provide feedback, ideas, or recommendations to enhance the portal, fostering user engagement and continuous enhancement of the platform's features and services | 0 - no, 1 - yes | 3 |
| (i) Portal sustainability and collaboration | (1) sustainability strategy | strategic plan that outlines how the portal aims to ensure the long-term availability, relevance, and impact of the data it hosts, often encompassing funding models, data governance, user engagement, and partnerships to support ongoing operations and growth | 0 - no, 1 - yes | 2 |
| | (2) performance insights dashboard | visual representations and numerical data that provide a quick and easily digestible overview of key performance indicators, allowing stakeholders to assess the effectiveness and progress of the portal | 0 - no, 1 - yes | 3 |
| | (3) regional governments collaboration mentions | information about cooperative efforts and partnerships with local governments or institutions | 0 - no, 1 - yes | 2 |
| | (4) international collaboration mentions | information about cooperative efforts and partnerships with international governments or institutions | 0 - no, 1 - yes | 3 |
| | (5) user satisfaction survey | assessment to collect feedback and opinions from users regarding their experiences and contentment with the services, data accessibility, and overall performance of the portal is conducted | 0 - no, 1 - yes | 3 |
| | (6) open source codebase | code is open source, and the link to the repository is publicly available on the portal | 0 - no, 1 - yes | 1 |

The overall portal score is determined by summing these multiplied values (see equation 1):

$$\gamma = \sum(x_l) * 1 + \sum(x_m) * 2 + \sum(x_h) * 3 \tag{1}$$

where $\gamma$ is the overall score, $x_l$, $x_m$, $x_h$ - is the score of the sub-dimension marked as *low*, *medium*, *high* importance, respectively.

Table 3. Weighing systems used in the existing research. *Asterisks (*) means multiple indices are mentioned in 1 study*

| Category | References |
|---|---|
| Equal weights for dimensions and aspects | (Lnenicka & Nikiforova, 2021; Zhu & Freeman, 2019; Máchová & Lnénicka, 2017), WJP Open Government Index (Susha, et al., 2015)* |
| Equal weights for dimensions, different for aspects | (Sisto et al., 2018; Raca et al., 2021), ODI Barometer, ePSI Scoreboard (Susha et al., 2015)* |
| Different weights for dimensions and aspects | (Carsaniga et al., 2022; Knippenberg, 2020; 2020; Herrera-Melo & González-Sanabria, 2020; Wang et al., 2018), Open Data Watch (Zuiderwijk et al., 2021)* |
| Equal weights for dimensions | Capgemini OD Economy (Susha et al., 2015)*, Open Government Data Report (Zuiderwijk et al., 2021)* |
| Different weights for dimensions | (Afful-Dadzie & Afful-Dadzie, 2017), OKNF OD Index (Zuiderwijk et al., 2021)* |
| Importance/priority factor involved | World Bank ODRA (Susha, et al., 2015)* |
| Aggregated results | (Alexopoulos et al., 2018; Nikiforova, 2021; Abella, Ortiz-De-Urbina Criado, & De-Pablos-Heredero, 2022) |

# 5 Application of the developed framework to 33 OGD portals: analysis of the results

The integrated OGD portal usability framework was applied to 33 EU and GCC OGD portals, calculating individual portal scores using Equation 1 for each dimension and summing these to produce a total score, where the maximum possible score is 177 points.

*France* leads the portal ranking (Figure 2), consistent with previous studies (e.g., (Carsaniga et al., 2022)). Although the French portal did not achieve the maximum score of 177, its score of 141 is notably high, placing it 19 points ahead of *Saudi Arabia* (122) in second place. These results are competitive, with an average portal score of 84.9, an EU-only average of 88.7, and a GCC average of 67.8. Eleven national OGD portals exceeded the 100-point threshold, while only four scored below 50, with Kuwait's OGD portal ranking last.

*Kuwait*'s low rank is likely due to the absence of a national open data portal, relying instead on a dedicated OGD section within the government's portal.

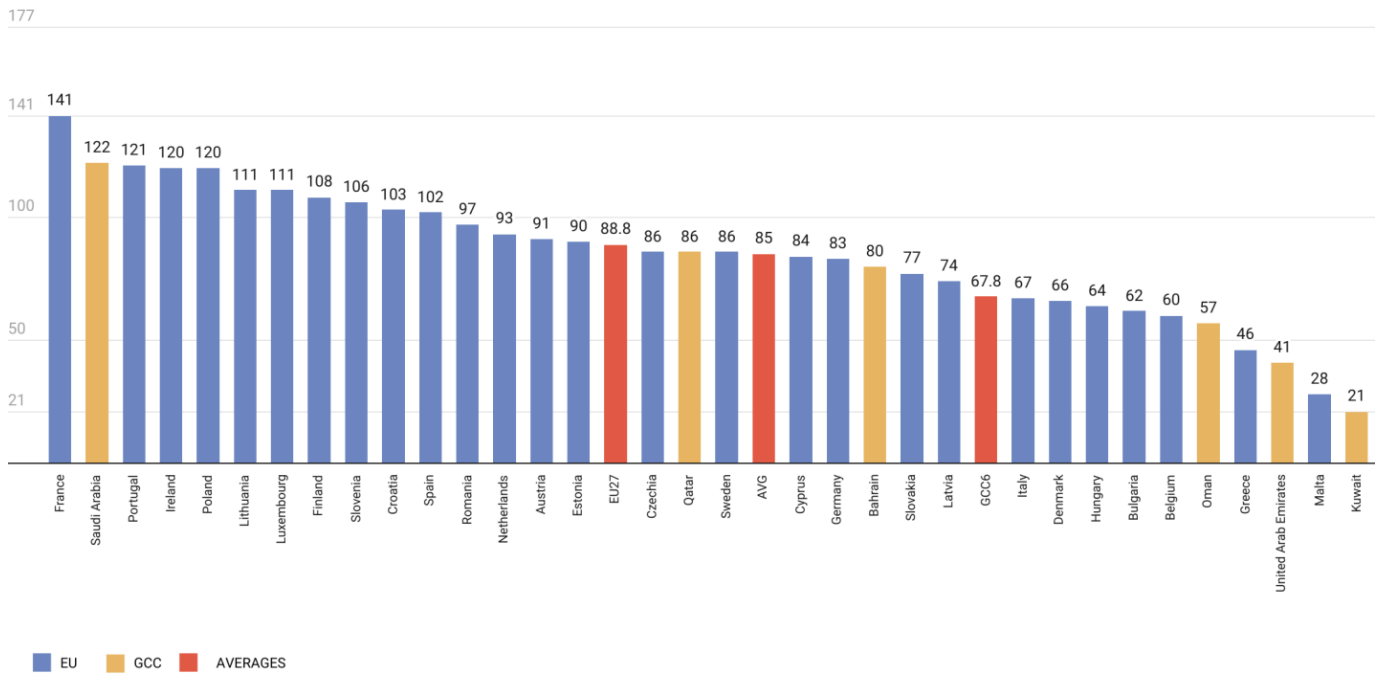

Figure 2. Portal ranking

Let us now discuss the results demonstrated by the selected portals by dimension.

## 5.1 Multilingualism

In the "Multilingualism" dimension, eight countries' portals received the maximum score, namely *Bahrain, Estonia, Ireland, Malta, Oman, Qatar, Saudi Arabia*, and the *United Arab Emirates* (Figure 3). Notably, five of these are GCC states, reflecting their commitment to cross-regional collaboration through extensive English language support.

Conversely, most EU portals scored minimally. The portals of *Germany*, *Hungary*, and *Italy* scored zero as they only support German, Hungarian, and Italian, respectively. In contrast, the *Irish* and *Maltese* portals scored highly because English is one of their official languages.

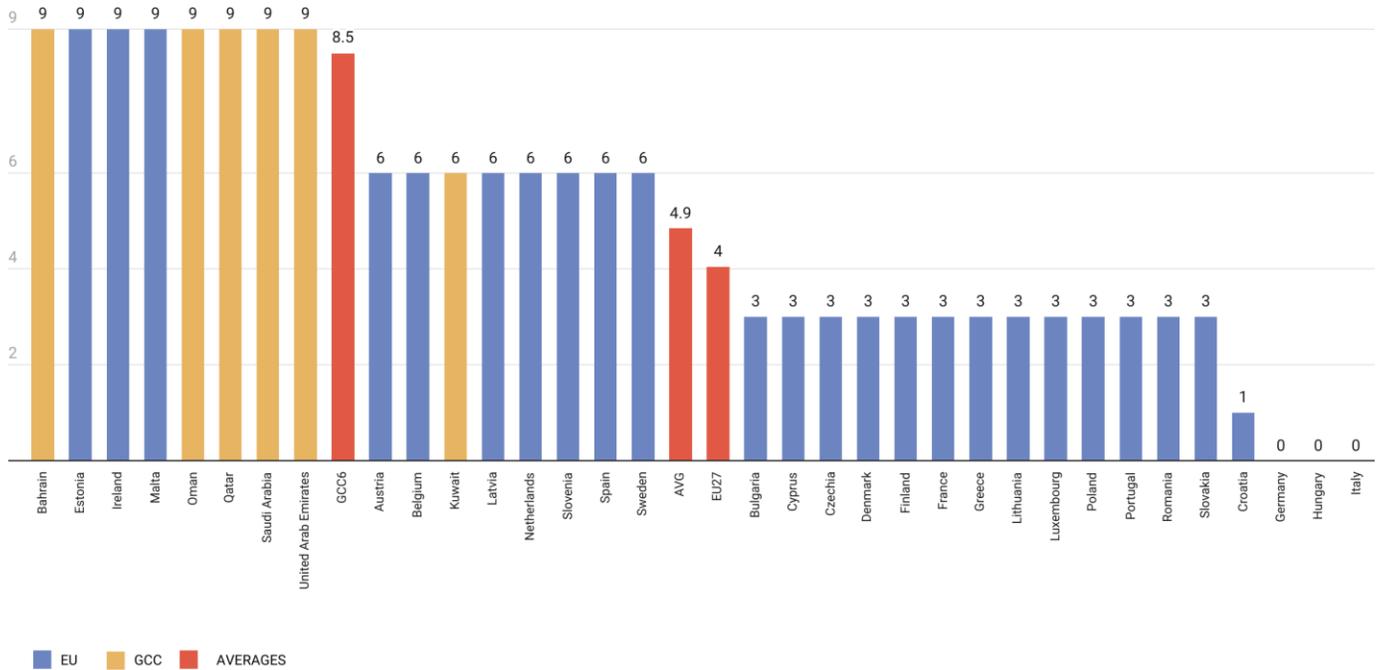

Figure 3. Ranking in "Multilingualism"

Several positive examples can be emphasized drawing from our sample and associated results. The *Estonian* portal demonstrates effective use of machine-translated metadata, translating metadata and informing users about potential low-quality translations (Figure 7 (a)). The *Austrian* portal, on the other hand, redirects users to the EU Open Data portal for machine-translated metadata in their selected language and allows publishers to optionally provide English titles and descriptions. The *Slovenian* portal employs the Google Translate plug-in for portal-wide translation, covering both content and user interface (Figure 7 (b)). However, this approach has the drawback of search functionality relying on the original Slovenian metadata.

Overall, multilingual support across the analyzed portals is **fairly basic** and needs improvement, especially in

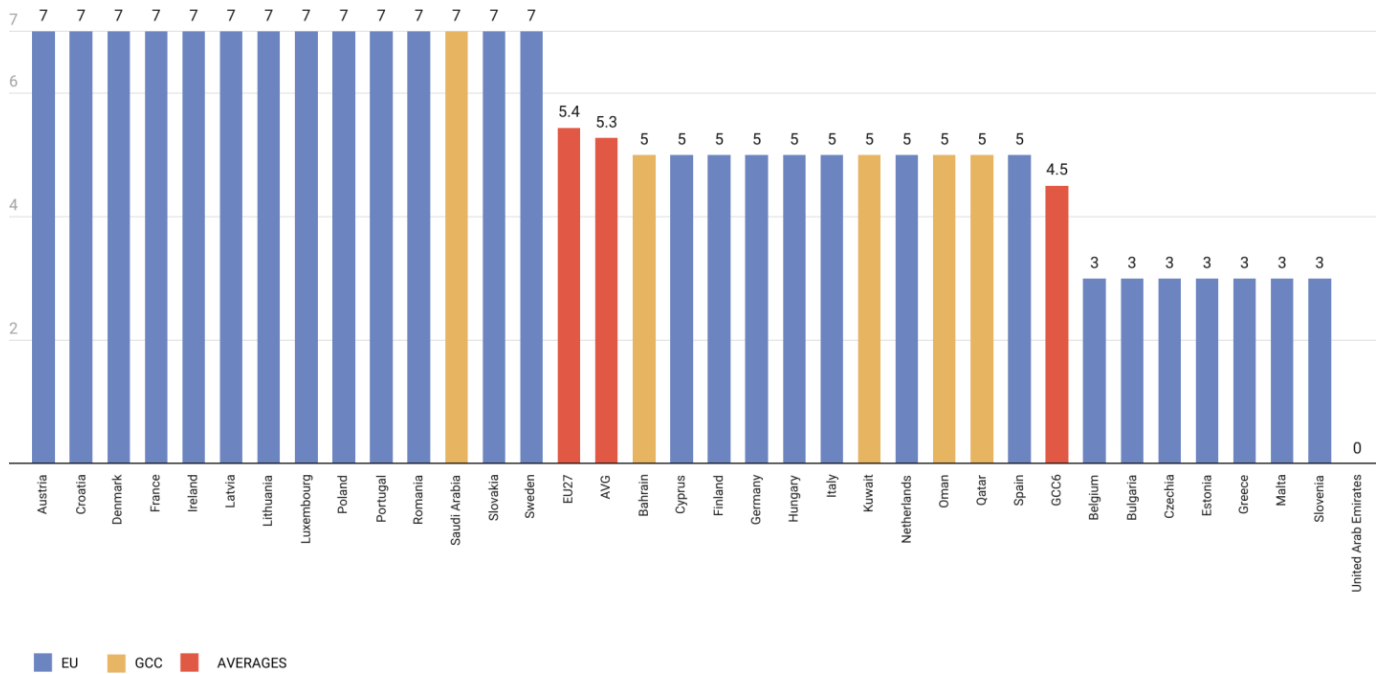

Figure 4. Examples of "Multilingualism" dimension features - machine-translated metadata and Google Translate plug-in in Estonian and Slovenian portals, respectively enabling users to search datasets in their preferred language.

## 5.2   Navigation

In the "Navigation" dimension, fourteen of the thirty-three countries' portals achieved the maximum score, namely *Austria, Croatia, Denmark, France, Ireland, Latvia, Lithuania, Luxembourg, Poland, Portugal, Romania, Saudi Arabia, Slovakia*, and *Sweden* (Figure 5). Among the top performers, only *Saudi Arabia* represents the GCC states. Conversely, the *United Arab Emirates* portal significantly impacted the regional average by failing to score points.

An interesting observation, although not factored into the final score, was the inclusion of a site map on the portals of the *Czech Republic, Germany, Poland*, and *Italy*.

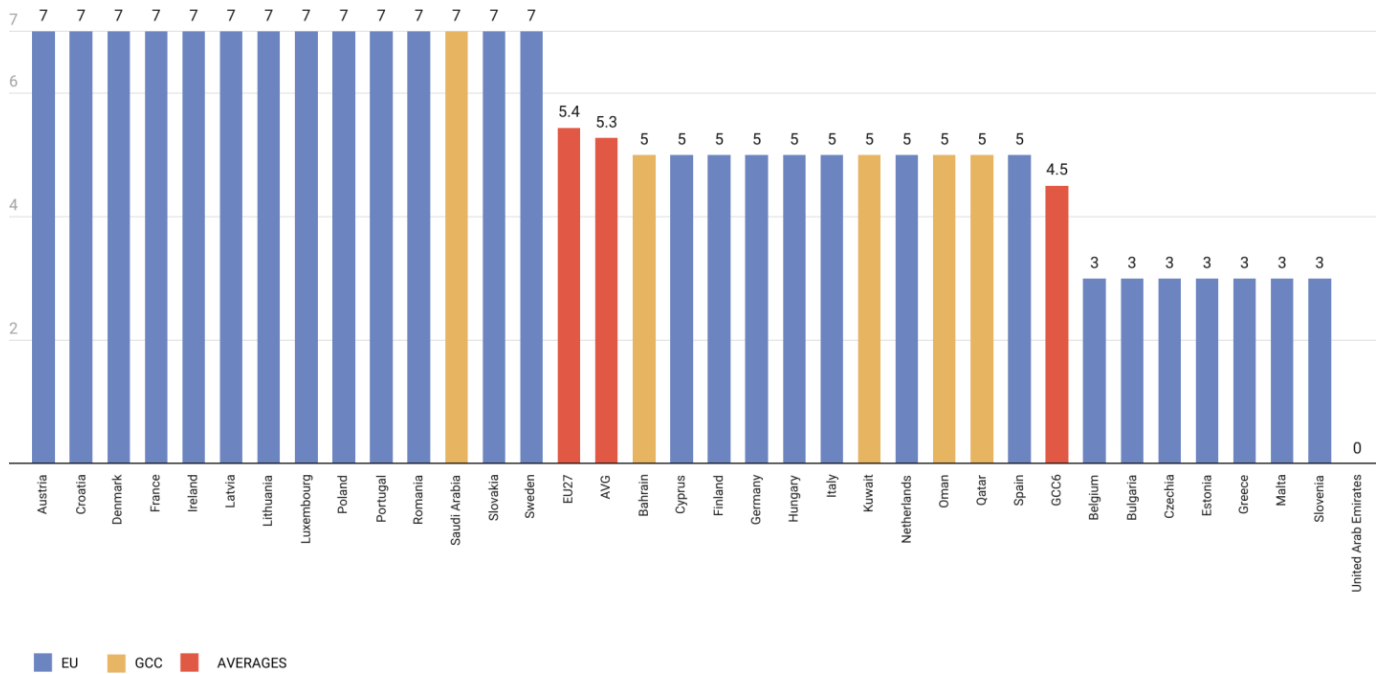

Figure 5. Ranking in "Navigation"

For instance, the *Polish* portal places all primary links prominently in the footer, facilitating easy access to each section. The *French* portal uses tabs filled with a balanced amount of information and interconnected pages to enhance navigation intuitiveness (see Figure 6).

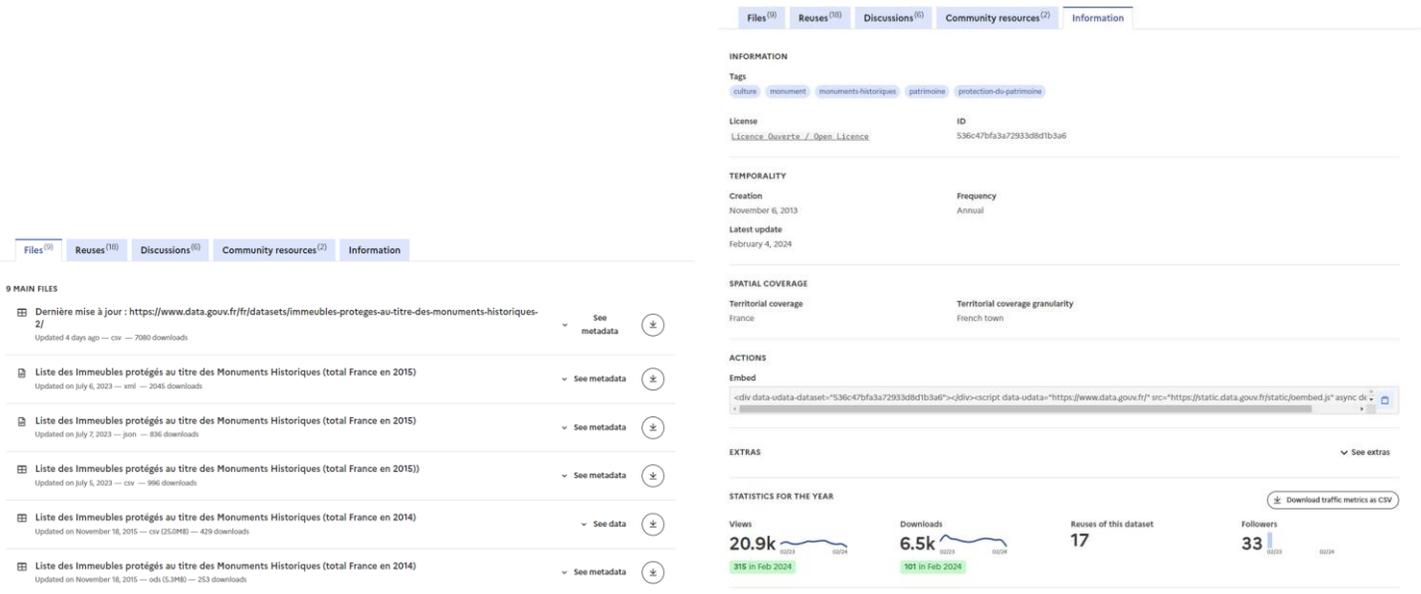

Figure 6. Tabs on the dataset page of the French portal assist in organizing data

However, the presence of navigation elements alone does not guarantee their intuitiveness or practicality. For example, breadcrumbs on the *Austrian* portal may not direct users to previously opened pages. Clicking on tabs on the dataset page of the *Bulgarian* portal redirects to new pages without a link back to the dataset's main page. Some portals, like the *Portuguese* portal, prioritize publisher information over dataset metadata on the dataset page, which may not be user-friendly, while the *UAE* portal hides the menu bar on dataset pages, making the content less consistent compared to other pages.

Overall, portals **typically provide essential navigation elements**, but **some lack consistency**. Additionally, in some cases, like the *Irish* portal, menu bar sections may be flattened or simplified, making navigation easier.

## 5.3 General performance

In the "General performance" dimension, fifteen countries' portals achieved the maximum score, namely *Austria, Belgium, Croatia, Finland, France, Italy, Latvia, Lithuania, Luxembourg, Netherlands, Poland, Portugal, Slovenia, Spain*, and *Sweden* (Figure 8). Notably, none of the GCC states reached the top.

However, some problems have been identified. In the *Czech* portal, a "delete catalog" (delete dataset) button is present on the dataset page, which is visible not only to the dataset owner (See Figure 9). It is unclear why regular users would have access to the button that leads to the instructions on creating a deletion request for the dataset. It is important to note that this feature is not connected to data error reporting, which sends reports to data curators. For most portals analyzed, the general performance is acceptable, with the portals of *Poland*, *Bahrain*, *Estonia*, and *France* serving as exemplary models of responsiveness and robustness.

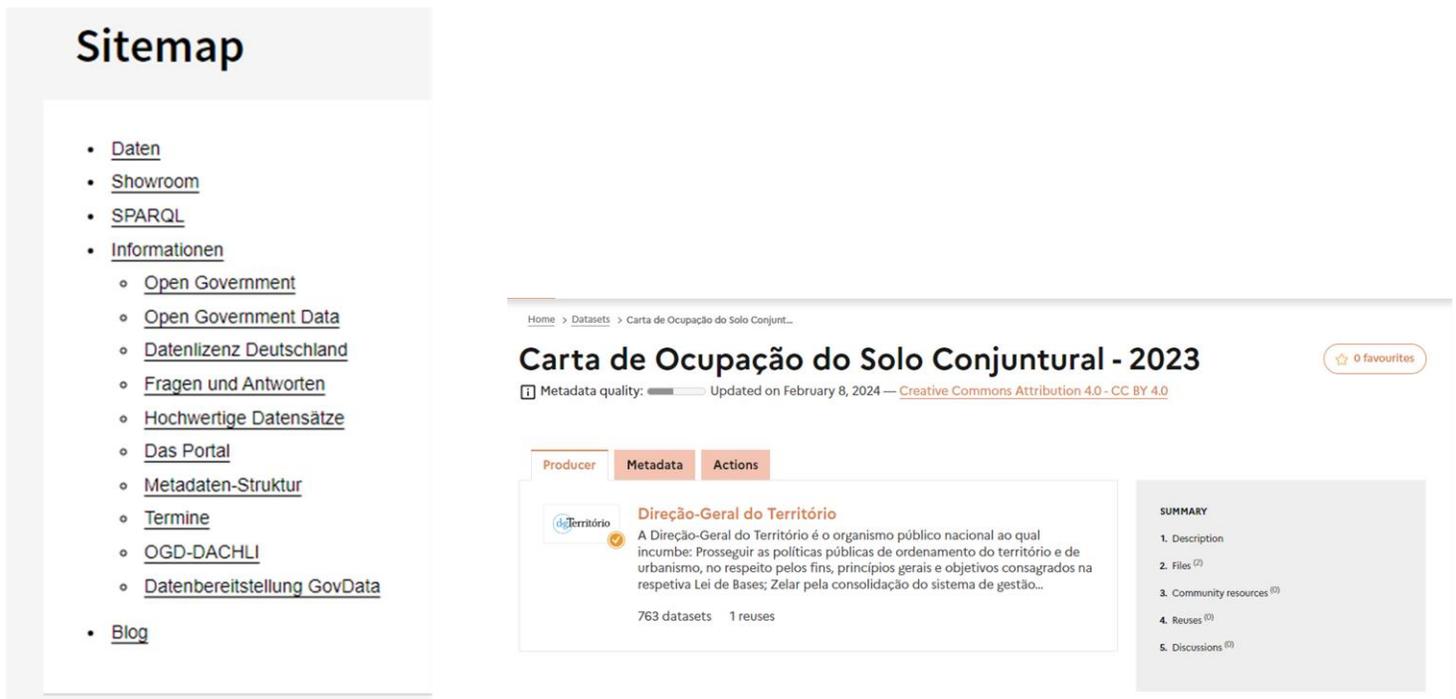

Figure 7. (a) The German portal's sitemap, (b) In the dataset page, the producer's information tab displayed by default instead of metadata information

The *Maltese* portal displays internal identification keys on its pages. The Omani portal has technical restrictions when downloading data, limiting users to 2 million rows per download. The *Irish* portal has issues with loading some dataset pages, particularly the last two pages of the oldest datasets. The *Greek* OGD portal requires registration to access data via API tokens, but users did not receive confirmation upon attempting to register.

Additionally, the unavailability of dataset resources (files) is common among portals. Only the portals of *Latvia*, *Luxembourg*, *Portugal*, *Qatar*, and *Saudi Arabia* had resources available for all sampled datasets. However, the dataset sample may not fully reflect the overall situation regarding resource availability.

Overall, portals could benefit from placing **more emphasis on assessing the quality of portal functionality**

and improving page loading speed, which was observed to be rather slow.

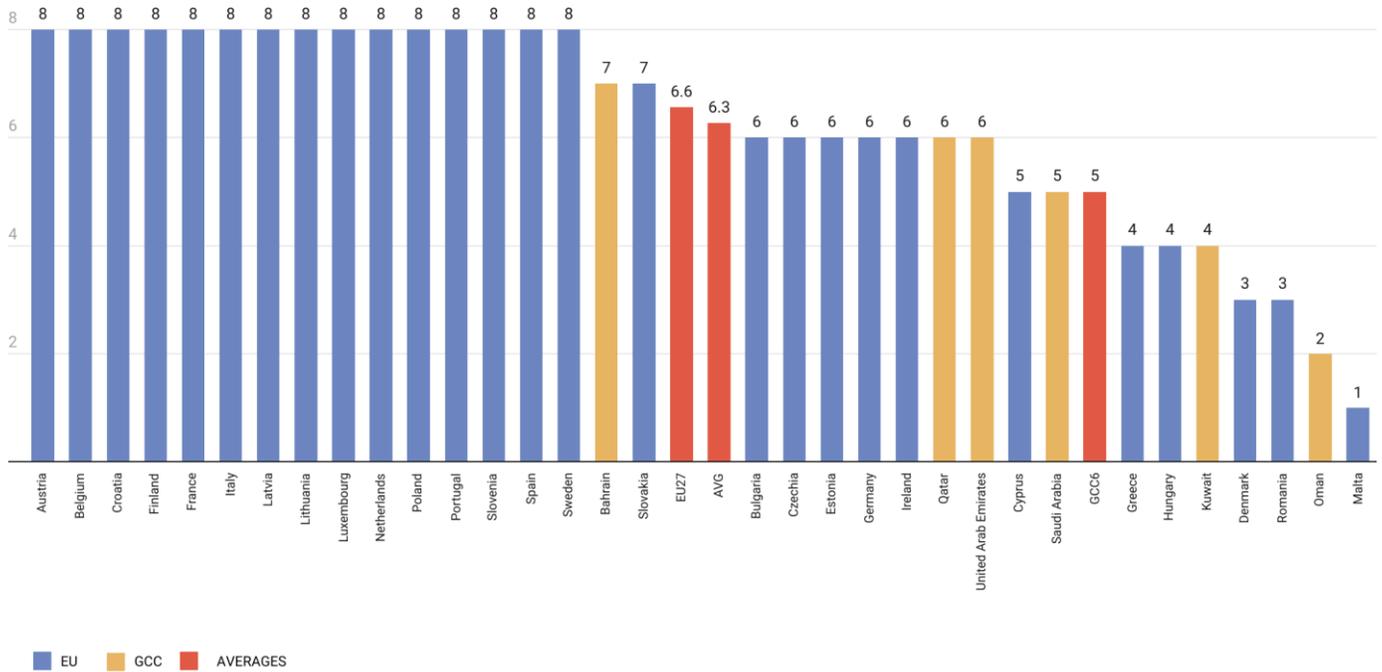

Figure 8. Ranking in "General performance"

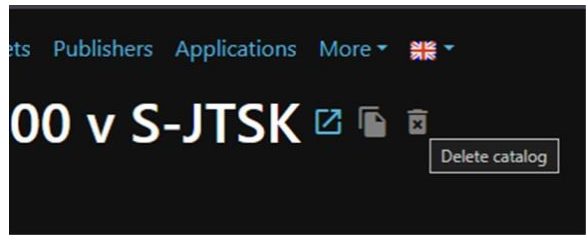

Figure 9. Czech portal: "delete catalog" button on the dataset page displayed for all users

## 5.4   Data understandability

In the "Data understandability" dimension, no portal achieved the maximum score of 26, as shown in Figure 10. The *Polish* portal attained the highest score, while the *Qatari*, *French*, and *Saudi Arabian* portals followed closely behind. Notably, among the top five performers, three are from GCC states. Conversely, seven portals—*Bulgarian*, *Estonian*, *Italian*, *Kuwaiti*, *Maltese*, *Swedish*, and *UAE*—received a score of zero.

The *Dutch* portal's impact section and success stories (also use-cases/showcases/re-uses) on the *French*, *Portuguese*, and *German* portals highlight applications and services built on open data, showcasing practical uses and benefits. Unfortunately, the latter is the most effective and most resource-consuming.

Promoting high-value datasets (HVD) is also crucial. This can take various forms, such as additional filtering criteria (seen in the portals of *Ireland*, *Lithuania*, *Poland*, and *Slovenia*), featured lists (as in the *Dutch* portal), or reports highlighting valuable datasets (as in the *French* portal). The *Czech* portal promotes HVD by holding publishers accountable through indicators and dashboards that measure dataset performance.

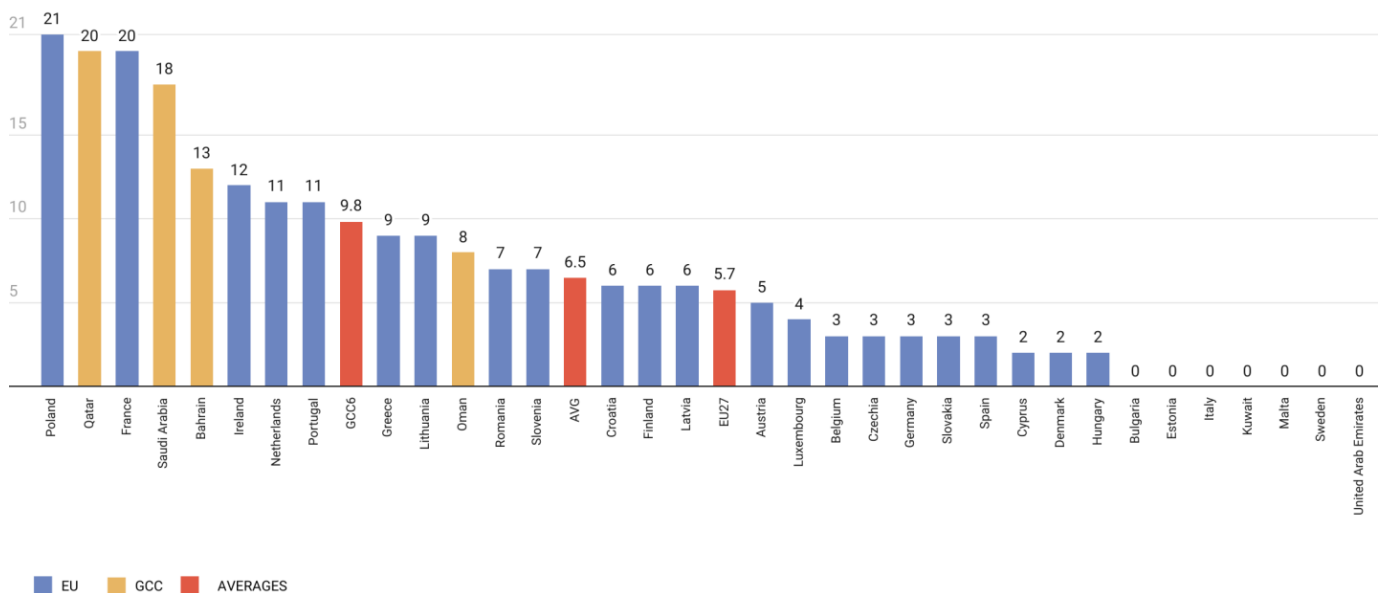

Figure 10. Ranking in "Data understandability"

This dimension receives **the lowest average scores**, indicating a need for portals to put **more effort into describing data use, its implications, and benefits**. Emphasis should be placed on **promoting high-value datasets** and introducing **visualization and analysis tools**. **Visual cues** can aid users in understanding the content. For example, the *French* portal provides insightful statistics and charts on the "info" tab of dataset pages However, these may go unnoticed without clear indicators that such charts are available.

## 5.5   Data Quality

In the "Data Quality" dimension, only the *French* portal attained the maximum score of 25 (Figure 11). The *Saudi Arabian* portal is the sole GCC representative in the top 10, ranking 9th alongside the *Slovakian* portal. The *Kuwaiti* portal, however, scored zero.

A common feature of Top performers in the data quality dimension - *France, Czechia, Slovenia, Portugal*, and *Croatia* - is the use of indicators that provide insights into the availability and accessibility of various dataset aspects, including resources, specifications, and update frequency accuracy. The Czech portal checks for personal data in datasets (Figure 13a). The French and Portuguese portals offer a dataset metadata quality indicator to inform users and publishers about missing or inaccurate metadata details (Figure 12).

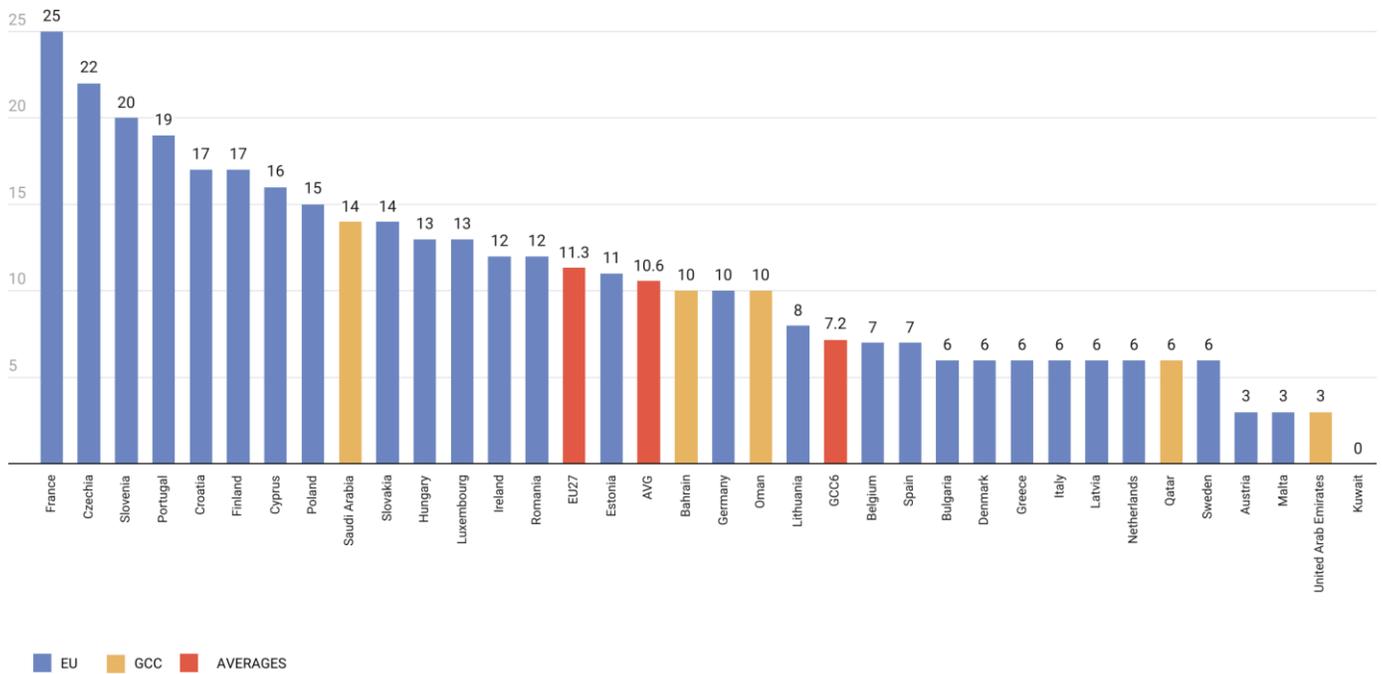

Figure 11. Ranking in "Data Quality"

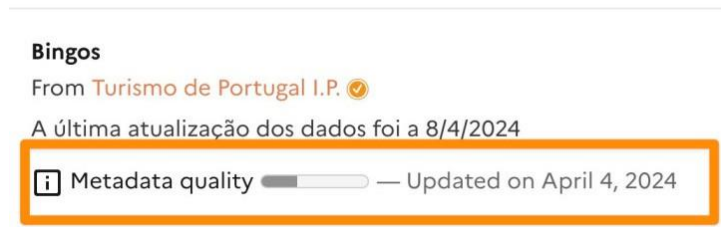

Figure 12. Portugal's portal dataset metadata quality indicator

The *Romanian*, *Slovenian*, and *Irish* portals use a 5-star scheme dataset openness rating, making it easier to determine the openness of datasets.

The *Bahrain* and *Qatar* portals provide extensive explanations of dataset schema. The *Qatari* portal allows catalog downloads in RDF format, while the *Polish* portal allows dataset metadata downloads in CSV and RDF formats.

The *Cyprus* portal effectively utilizes the temporal coverage parameter, though its spatial parameter usage may be subject to debate since "Cyprus" value is used for all datasets without more granular division. The *Finnish* portal integrates spatial coverage into interactive visualizations and catalog filtering (see Figure 13b).

The *Romanian* portal has a resource availability indicator, but it is not prominently placed on the resource page (located at the bottom).

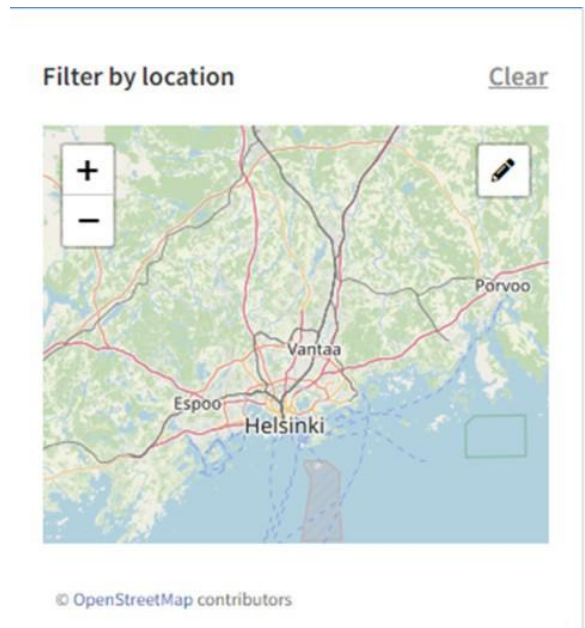

(b) The Finnish portal integrates spatial coverage into catalog filtering

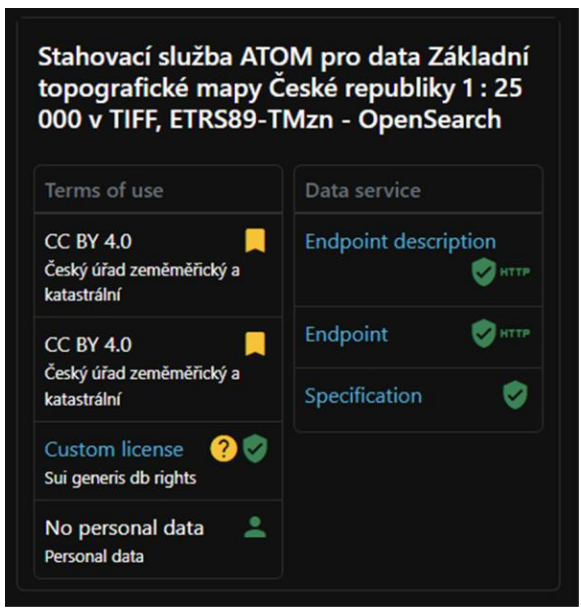

(a) The Czech portal checks whether datasets contain personal data

Figure 13. Examples of "Data Quality" features

In general, portals should **improve their metadata collection and provision standards** to ensure richer and more consistent data across datasets. The introduction of **quality indicators** is a positive trend that should become more widespread. Additionally, enabling **dataset metadata downloads** would be a beneficial feature.

## 5.6   Data findability

In the "Data Findability" dimension, no portal attained the maximum score of 38, as shown in Figure 14. Strong performers with scores over 30 include portals from *Ireland*, *Luxembourg*, *France*, *Austria*, and *Poland*. The *Saudi Arabian* portal is the sole GCC representative among the top 10 performers, while the *Kuwaiti* portal received a score of zero.

The *Irish* portal stands out for providing advanced search capabilities, while the *Portuguese* and *Swedish* portals offer users search tips (Figure 15). The *Finnish* portal allows users to search data within the selected region on the map.

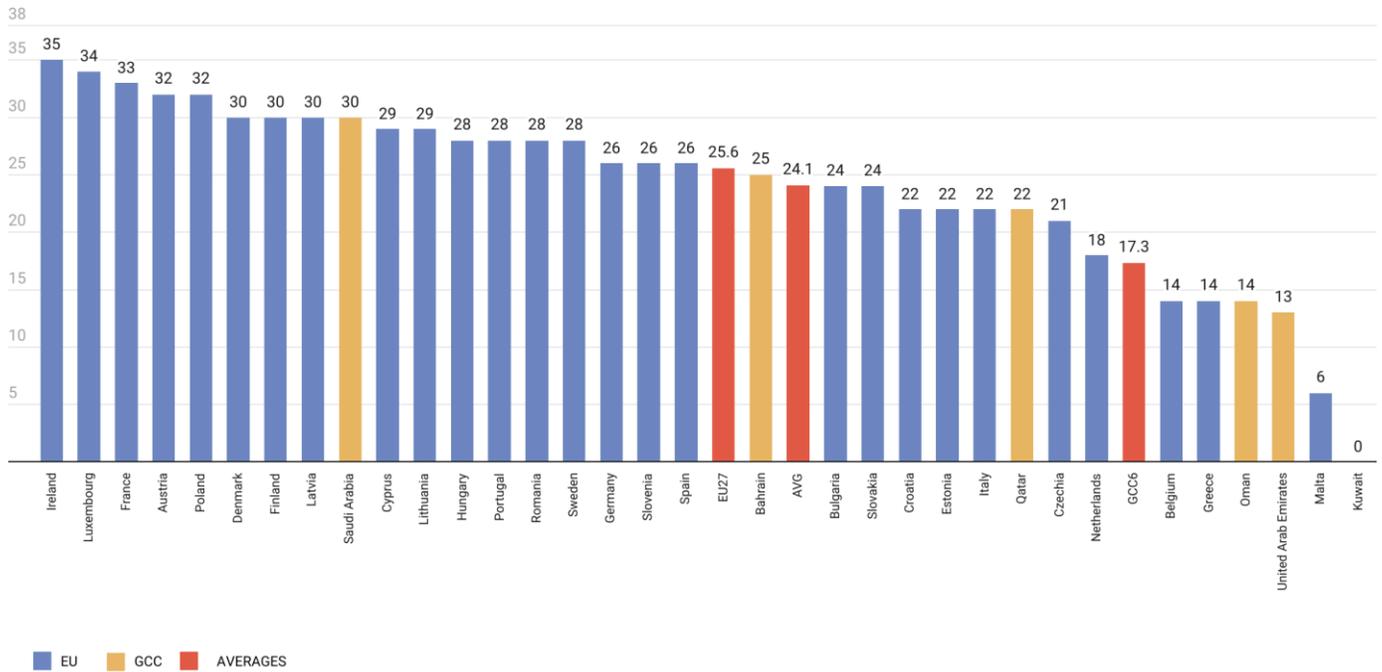

Figure 14. Ranking in "Data findability"

The *French, Italian, Dutch, Polish, and Luxembourgish* portals provide exemplary illustrations of how to implement *"featured topics"* sections. These topics may be general, but for some portals they are quite specific. For example, the *Polish* portal displays a collection of datasets related to Ukraine, while the *French* portal provides a rich list of topic-specific featured datasets on topics such as energy, education, culture, COVID-19 etc. Related datasets are displayed on the dataset page in the *French* and *Dutch* portals. Unfortunately, the likeness relation/similarity rate is not shown.

The *French*, *Italian*, *Dutch*, *Polish*, and *Luxembourgish* portals offer exemplary implementations of "featured topics" sections. While these topics may be general, some portals feature specific topics. For example, the *Polish* portal displays a collection of datasets related to Ukraine, while the *French* portal provides a rich list of topic-specific featured datasets covering areas such as energy, education, culture, and COVID-19. Related datasets are displayed on the dataset page in the *French* and *Dutch* portals. Unfortunately, the likeness relation/similarity rate is not displayed in these portals.

The *Omani* and *Greek* portals require authentication to access their data through the API, contradicting the openness principle, especially when it's the only way to download data from these portals. Similarly, the *Danish* portal exposes endpoints that return no content.

Figure 15. Search tips in the Swedish portal catalog

Overall, portals **perform adequately** in this dimension. However, there's room for **improvement in exposing API/GraphQL endpoints**, ensuring **content accessibility**, **establishing connections between datasets based on similarity** to facilitate promotion, and highlighting **featured topics**.

## 5.7   Public engagement

In the "Public Engagement" dimension, there's a notable gap between the best performer, represented by the *Lithuanian* portal, and the maximum score, as shown in Figure 16. Among the GCC portals, only the *Saudi Arabian* portal surpasses the average, while the *UAE* and *Malta* portals received a score of zero.

The *Lithuanian* portal excels in the public engagement dimension with its lively news section containing an abundance of articles and event announcements (See Figure 17a). The *Spanish* and *Croatian* portals stand out for their rich report-tracking features, displaying a list of reports and their status (See Figure 17b). Additionally, several portals, including *Spanish*, *Estonian*, *Irish*, *Lithuanian*, and *Croatian*, allow users to report issues, provide suggestions for improvement, and exchange information regarding reuses and initiatives.

Some portals incorporate video content, such as interviews and educational material (the *Polish*, *Czech*, and *Spanish* portals). The Polish portal offers users the option to receive dataset or search result updates via notifications. Noteworthy examples of community-sourced datasets, maintained not by administrative or governmental institutions, have been observed on portals like *France* and *Finland*.

On average, portals **perform relatively low** in this dimension. Few portals introduce advanced **gamification elements**, and **personalization options** are often limited. Many portals lack **a catalog of previously reported issues** or the **ability to upload reuses**. Additionally, **social media accounts** associated with these portals are rarely active.

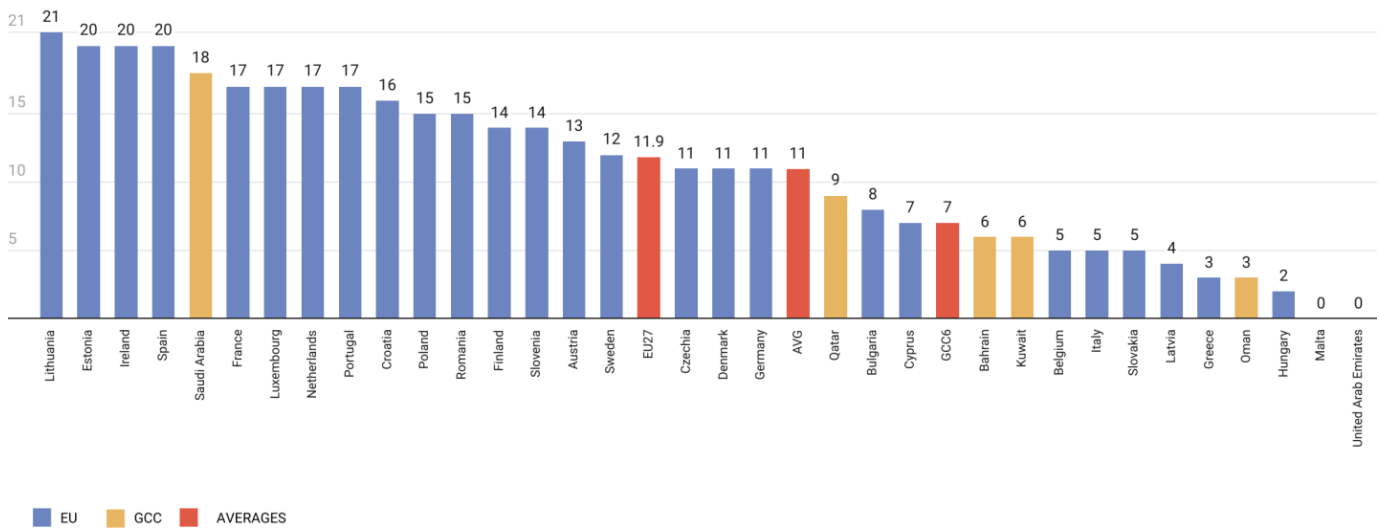

Figure 16. Ranking in "Public engagement"

| (a) News page in the Lithuanian portal | (b) Request tracking chart in the Spanish portal |

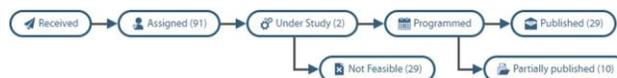

(a) News page in the Lithuanian portal    (b) Request tracking chart in the Spanish portal

Figure 17. Examples of "Public engagement" features

## 5.8   Feedback mechanisms and service quality

In the "Feedback mechanisms and service quality" dimension, *Spain, Croatia, Lithuania, France, Portugal* and *Slovenia* are the top performers in the *feedback mechanisms and service quality* dimension, as shown in Figure 18.

However, none of them reached the maximum score (18). Being the only one from the GCC, the *Saudi Arabian* portal once more ranked among the top ten. The portals from *Bahrain, Denmark, Greece, Kuwait and Malta* received a score of zero.

In the "Feedback Mechanisms and Service Quality" dimension, *Spain*, *Croatia*, *Lithuania*, *France*, *Portugal*, and *Slovenia* emerge as the top performers, although none of them reached the maximum score ( Figure 18). Notably, the *Saudi Arabian* portal ranks among the top ten. Conversely, the portals from *Bahrain*, *Denmark*, *Greece*, *Kuwait*, and *Malta* received a score of zero.

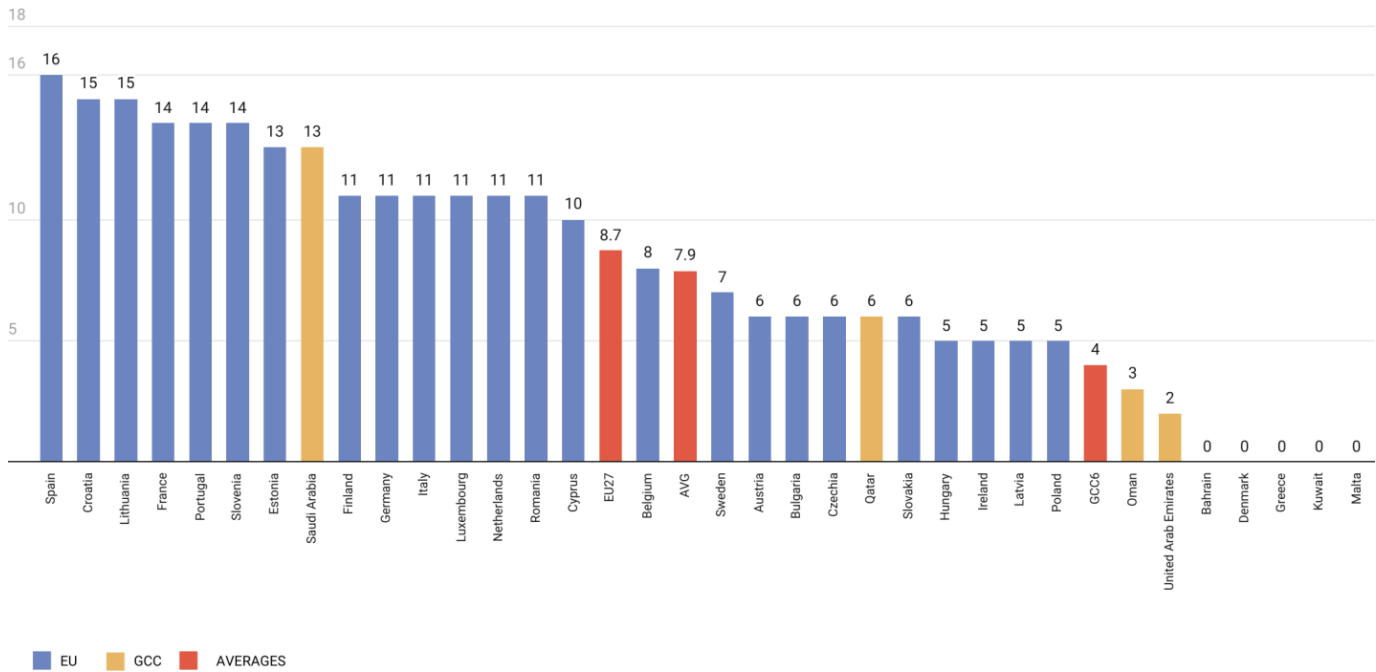

Figure 18. Ranking in "Feedback mechanisms and service quality"

A common trend observed is the inclusion of a comment section on the dataset page in portals of *France*, *Croatia*, *Lithuania*, and *Luxembourg* (see Figure 19a). The usefulness of this feature can be verified by observing the lively discussion in the corresponding section, where, however, the participation of both parties, i.e., not only the user but also the publisher, is important.

Several portals implement dataset usefulness assessment mechanisms, ranging from upvoting/liking (e.g., the *Croatia, the Netherlands*) to scoring scales (e.g., *Estonia*) or subscribing/following datasets or pages ((e.g., the *France, Portugal, Latvia*).

Many portals (e.g., the *French, Portuguese, Austrian, Czech* ones) provide users with - potential publishers and regular users - *guides and manuals*. However, they are often very technical and are unlikely to be understandable to a lay user or tailored to data publishers. Although there are examples (e.g., the *Irish, Dutch, and Czech* portals) where some manuals are tailored toward lay citizens. In terms of guides and manuals, while many portals offer them, they often tend to be overly technical and are unlikely be understandable to a lay user or tailored to data publishers. Some portals, such as the *Irish*, *Dutch*, and *Czech* ones, however, provide manuals tailored toward lay citizens. Additionally, features like a virtual tour, as seen in the *Polish* portal, can greatly assist new users (Figure 19b). The *Saudi Arabian* portal stands out for offering a diverse range of communication channels, including mail, contact forms, addresses, dataset suggestions or requests, and complaint forms.

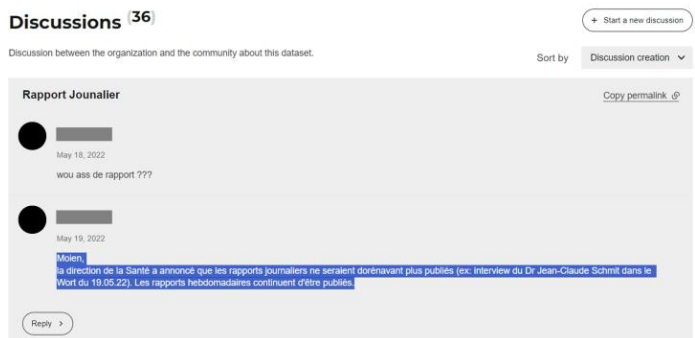
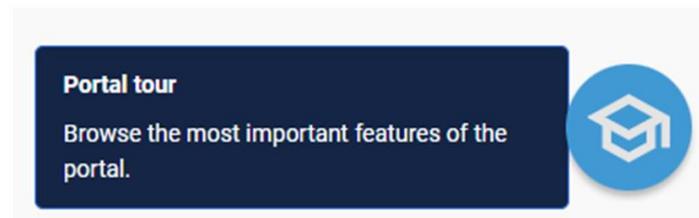

(a) Lively comment section in the Luxembourgish portal (comments are anonymized)

(b) Polish portal's virtual tour button

Figure 19. Examples of "Public engagement" features

In general, portals should continue improving their service quality and feedback functions. Portals should create and maintain comment sections to foster communication between publishers and users. Overall, there's **room for improvement** in service quality and feedback mechanisms. Portals should **prioritize creating and maintaining comment sections** to facilitate communication. **Manuals and documentation should be more user-friendly** and diverse in content (e.g., the *Austrian, Greek, Bulgarian, and Bahraini* portals) have no communication with the support service or it only relies on writing emails). Additionally, the adoption of **various communication channels can enhance user support and engagement**, where although using online chat for customer service in the *Belgian* portal is uncommon, other portals may find it advantageous to adopt this practice.

## 5.9   Portal sustainability and collaboration

In the "Portal Sustainability and Collaboration" dimension, six portals achieved the maximum score, namely the *Finnish*, *French*, *Irish*, *Luxembourgish*, *Polish*, and *Portuguese* portals (Figure 20). Notably, no portals from the GCC states are among the top 10 performers, and only the *Kuwaiti* portal hasn't scored.

Portals like the *Polish* portal exhibit a clear sustainability strategy, evident from the provision of the *Open Data Programme for 2021-2027* in the useful material section. On the other hand, the *French* portal tracks data releases monthly, highlighting released datasets and reuses (see Figure 21a). Additionally, the *German* portal celebrated its 10th anniversary in 2023 by sharing past milestones in a blog section.

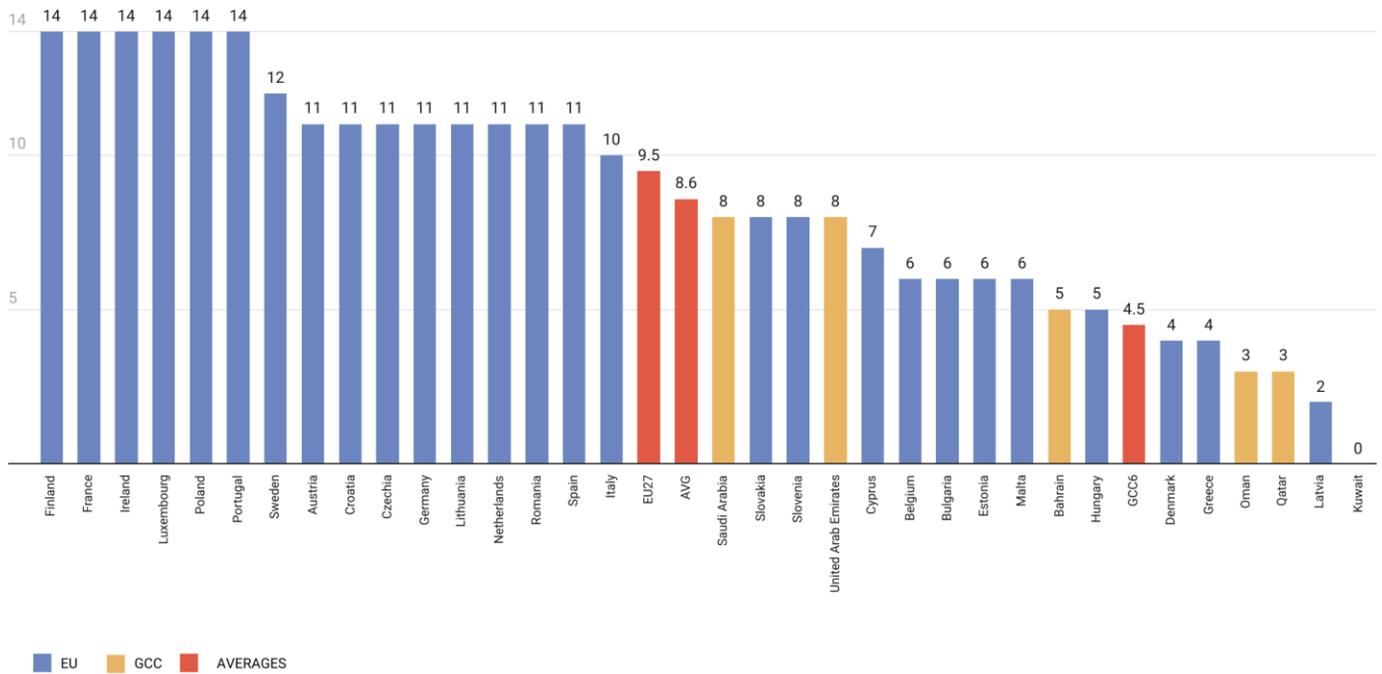

Figure 20. Ranking in "Portal sustainability and collaboration"

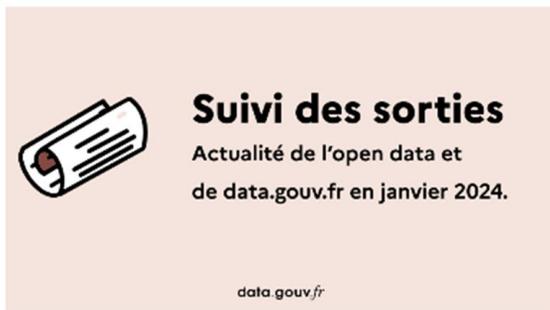

(a) French portal release tracker (News section)

(b) The Finnish portal user satisfaction survey

Figure 21. Examples of "Portal sustainability and collaboration" features

Table 4. Red cluster

| Cluster appearance | Countries |
|---|---|
| Both in K-means and hierarchical clusters | Bahrain, Greece, Kuwait, Malta, Oman, Qatar, United Arab Emirates |
| Only in K-means | Denmark, Hungary, Latvia |

Comprehensive user satisfaction surveys are rare but exemplified by portals like the *Finnish* and *French* ones (See Figure 21b). Instant satisfaction surveys can also be found on many pages of the *Saudi Arabian* portal.

The dimension of international collaboration is demonstrated by the *Germany-Austria-Switzerland-Liechtenstein cooperation* highlighted on the portal's strategy pages. The *Irish* portal promotes the regional OGD portal of *Northern Ireland*, while the *Qatari* portal allows users to build a map based on data connected to EU states and Israel.

Interestingly, the source code of the portals was found in associated repositories for 19 of the 33 analyzed portals, but few of the portals post links to the repository on the portal itself, where we found them through the Google then as an additional step of our analysis. GitHub is the platform of choice for the OGD portals to host their repositories.

In summary, while portals **generally perform adequately** in this dimension, there's room for improvement, particularly in conducting **more user satisfaction surveys**, defining **clear strategies**, **sharing reports**, and **tracking the release of new artifacts**. Emphasizing the benefits of **portal partnerships** can promote collaboration within and across regions, fostering a more comprehensive open data ecosystem.

## 5.10 Cluster analysis

To gain a deeper understanding of the relationships and similarities between various portals based on their performance metrics, clustering analysis was performed with two types of clustering analysis - K-means clustering, and hierarchical clustering - performed to group portals based on their sub-dimension performance.

The optimal number of clusters for K-means clustering was determined using the Elbow method, which indicated that partitioning the portals into four clusters would provide more distinct insights. Similarly, hierarchical clustering was performed with the selected number of clusters (4) determined by analyzing the dendrogram (see Figure 22). To refer to the merged clusters from both clustering types, color names were assigned: red, yellow, blue, and green. Tables 4, 5, 6, and 7 show the composition of each cluster.

The red cluster performs the worst in all dimensions, except for "Multilingualism" (best score among all clusters) and "Data Understandability," where it excels. The average score in *"Multilingualism"* is objectively high (7.01 out of 9), with a substantial gap separating the first and second places (5.39). However, it struggles in "Data Quality," "Data Findability," "Public Engagement," "Feedback Mechanisms & Service Quality," and "Portal Sustainability & Collaboration." Scores from those dimensions are more than twice as low as the maximum score. The portals of this cluster exhibit exemplary approaches relating to the *"Multilingualism" and "Data understandability"* dimensions.

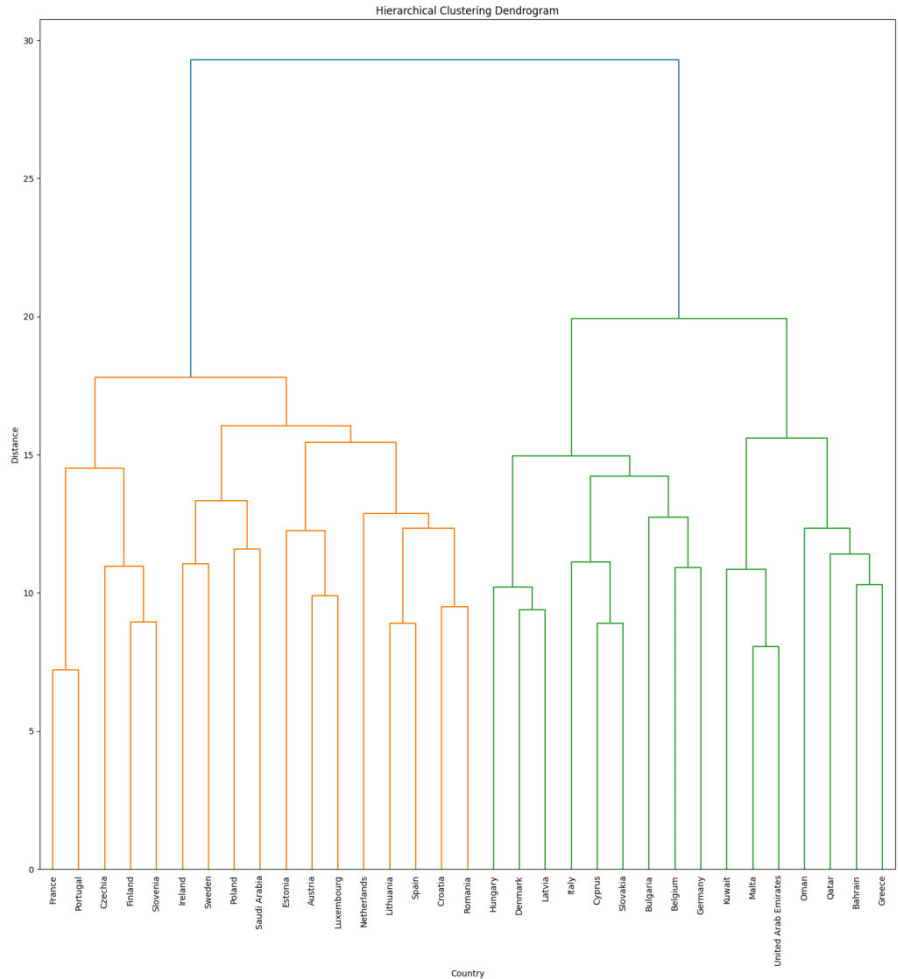

Figure 22. The dendrogram of the linkage matrix for hierarchical clustering

The yellow cluster ranks third in numerous dimensions and performs particularly poorly in "Multilingualism" and "Data Understandability" with just 2.17 out of 26 points for the later. Additionally, the dimensions where the cluster scored less than 50% of the maximum score are *"Data quality", "Public engagement", and "Feedback mechanisms & service quality"*. In other dimensions, the performance is rather acceptable.

The blue cluster is the second-best performing in most dimensions, excelling in "Navigation" and "Public Engagement." However, it faces challenges in "Data Understandability" and "Data Quality." For most dimensions ("Navigation", "General performance", "Data findability", "Public engagement", "Feedback mechanisms & service quality", "Portal sustainability & collaboration"), the score gap between the blue and green clusters is around 1-2 points. The portals of this cluster exhibit exemplary approaches in "Navigation" and "Public engagement" dimensions.

Table 5. Yellow cluster

| Cluster appearance | Countries |
| --- | --- |
| Both in K-means and hierarchical clusters | Belgium, Bulgaria, Cyprus, Germany, Italy, Slovakia |
| Only in K-means | Czechia |
| Only in hierarchical | Denmark, Hungary, Latvia |

Table 6. Blue cluster

| Cluster appearance | Countries |
| --- | --- |
| Both in K-means and hierarchical clusters | Austria, Croatia, Estonia, Ireland, Lithuania, Netherlands, Romania, Spain, Sweden |
| Only in hierarchical | Luxembourg, Poland, Saudi Arabia |

The green cluster is the best-performing overall, however, it lags behind other clusters in "Navigation", "Multilingualism", "Public engagement" with average results in later two and "Data understandability" (less than 50% of the maximum score). Although the cluster is the best-performing, only the scores in dimensions *"Navigation", "General performance", and "Portal sustainability & collaboration"* are nearly at their maximum. The portals of this cluster exhibit exemplary approaches in "Data findability", "Portal sustainability & collaboration" dimensions.

Table 7. Green cluster

| Cluster appearance | Countries |
| --- | --- |
| Both in K-means and hierarchical clusters | Finland, France, Portugal, Slovenia |
| Only in K-means | Luxembourg, Poland, Saudi Arabia |
| Only in hierarchical | Czechia |

The charts in Figure 23 show the merged clusters' performance comparison across nine dimensions.

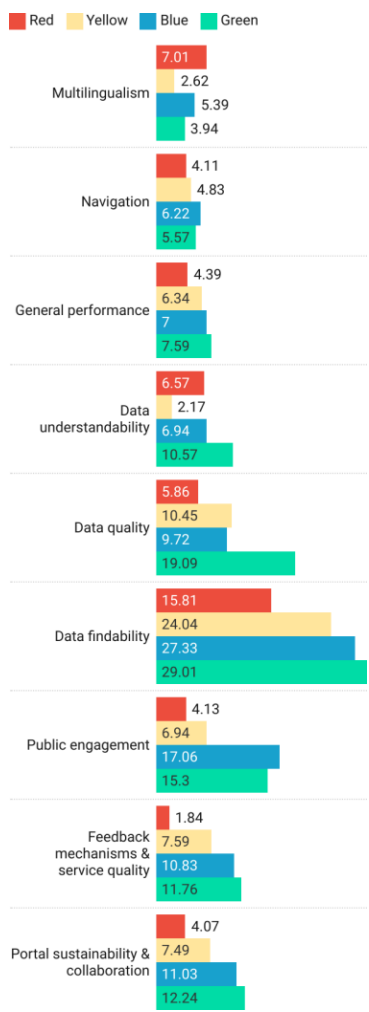

Figure 23. Cluster dimensional comparison

In summary, while each cluster exhibits strengths and weaknesses across various dimensions, the green cluster emerges as the most robust performer overall, demonstrating exemplary approaches in "Data Findability" and "Portal Sustainability & Collaboration."

# 6   Discussion

The integrated OGD portal usability framework serves as an instrument for evaluating various aspects and the overall performance of OGD portals, focusing on three key areas: (1) inclusivity, ensuring accessibility to a wide range of users; (2) support and facilitation of user collaboration and active participation; and (3) facilitation of data exploration and understanding. 21 out of 33 portals (18 EU, 3 GCC) exhibited no zero scores in any dimension, indicating that portal providers recognize the importance of implementing essential open data portal features, such as providing non-proprietary machine-readable data, as well as features that enhance user experience, attract more users, but also features that will attract more users to the portal, improve experience making them more user-friendly, and encourage repeated visits.

Among the 12 portals that have zero scores in certain dimensions, only four have multiple zero scores. This suggests that critical issues are generally confined to specific dimensions rather than being widespread. The dimension most affected by zero scores is "Data Understandability," with seven instances, followed by "Feedback Mechanisms and Service Quality" with five, "Multilingualism" with three, and "Public Engagement" with two. These dimensions, along with "Data Quality," are areas where portals tend to perform the worst.

Comparing the current results with existing studies ((Máchová et al., 2018; Carsaniga et al., 2022; Alexopoulos et al., 2018; Afful-Dadzie, 2017; Matheus et al., 2021; Zuiderwijk et al., 2014)) is challenging due to differences in dimensions, sub-dimensions, weighting systems, types of portals evaluated (city, regional, national), and the countries of origin. For example, local (city and regional) and national levels may not correspond directly in performance evaluations. Despite these differences, some comparisons can still be made. The ODM report series (Page et al., 2023; Carsaniga et al., 2022) is probably feature-wise the most up-to-date, although it is limited to EU rankings. While the dimensions differ from the present study's framework, 27 out of 33 portals assessed here are also evaluated in the ODM reports. In the Open Data Portal dimension, the top-5 EU portals in 2023 are from Poland, Estonia, Ireland, France, and Spain, whereas in 2022 they were from France, Poland, Ireland, Slovenia, and Cyprus. The current study's top-5 EU portals include France, Portugal, Ireland, Poland, and Lithuania. This comparison indicates that the current framework effectively identifies top performers across various dimensions. However, there are differences when discussing medium and low performers. For instance, Croatia and Czechia are among the worst in the ODM reports, but in the current study, Czechia is closer to the EU average, and Croatia is the 9th best performer among EU states. This discrepancy may be due to the more granular sub-dimensions in the current framework, providing a more detailed assessment that doesn't factor into the ODM framework. Alternatively, the difference may be explained by the fact that ODM reports are based at least partially on self-evaluation reports. This could lead to discrepancies in data due to two primary reasons - (1) there may be an incentive for those completing self-evaluations to report more favorable data to appear better compared to their peers, resulting in some data being unreliable as it may not accurately reflect the true state of the portals, (2) the data collected through self-evaluation might not always align with the actual conditions of the portals, where both a non-existent positive result might be reported, and possibly more common, existing positive results may not be reported due to a lack of awareness or knowledge by the individual responsible for providing the data (Lnenicka et al., 2024b; Bannister, 2007). These factors highlight the potential limitations of relying on self-evaluation reports and underscore the importance of independent assessments to provide a more accurate and comprehensive evaluation of OGD portals. Finally, in contrast to the ODM framework, which prioritizes portal-centricity, the framework introduced in this study focuses on users, which affects the final portal ranking.

Another study (Máchová & Lnénicka, 2017), employing the framework on top of which the framework for the present study is partially built, ranks Austria, France, and Croatia among the top performers. Although Austria did not rank in the top-10 in the current study, all three portals are high performers. A persistent criticism is that the quality of open data portals is influenced by the version of the data management system. The Portuguese portal, using the *udata* platform (Etalab, 2024) maintained by the French public agency Etalab, shares many features with the French portal, contributing to its high score. This plays a clear role in the high score of these portals.

A revised version of the above framework, the transparency-by-design framework (Lnenicka & Nikiforova, 2021), has been used in (Lnenicka, Nikiforova, et al., 2022) to test the maturity of transparency of smart city portals. Although this framework focuses more on transparency rather than usability, missing some current trends in the resilience and sustainability aspects of OGD portals, it also propagates the idea that portals should be adapted to users spanning from those with a very limited set of skills and low digital literacy to advanced users/experts; they should facilitate data exploration and encourage the re-use of the datasets (Lnenicka, Nikiforova, et al., 2022). That and the current study identify concerns regarding data quality and the absence of data quality monitoring.

A revised version of the above transparency-by-design framework (Lnenicka & Nikiforova, 2021) has been used to test smart city portal transparency in (Lnenicka, Nikiforova, et al., 2022). While it focuses more on transparency than usability, missing some current trends in resilience and sustainability, it supports the idea that portals should be user-friendly for both novices and experts, facilitating data exploration and encouraging dataset reuse. Both this and the current study identify concerns regarding data quality and the absence of data quality monitoring.

A study (Nikiforova, 2021) overlaps with the current study in tested portals but excludes Qatari, Saudi Arabian, and Kuwaiti portals. It shows that France, the Netherlands, Luxembourg, Estonia, and Portugal are top performers in real-time data. The current framework partially covers these aspects in the "Data Quality" dimension. In this study, France achieved the highest possible score in that dimension, Luxembourg and Estonia scored around the average, and the Netherlands performed poorly. These variations can be attributed to differences in assessment criteria and ongoing portal changes.

Comparing the findings of the present and previous research, the current framework effectively identifies high-performing portals and sheds light on additional features that could enhance portal usability. The findings highlight the need for better data quality metrics and improved communication channels between users and portal representatives. Additionally, the total absence of advanced gamification elements in portals, identified in previous research, suggests that incorporating such elements could encourage more frequent user engagement (Simonofski et al., 2022). Aligning with previous studies (Elsawy & Shehata, 2023; Nikiforova & McBride, 2021; Alexopoulos et al., 2018; Nikiforova et al., 2023; Chokki, n.d.; Benmohamed et al., 2024), this study confirms the need for enhanced English language support in the user interface, data search capabilities, and the promotion of high-value datasets (HVD), as well as introducing means to visualize and analyze datasets.

## 6.1 Recommendations

The results of the qualitative and quantitative analyses conducted in this study, along with the identification of best practices from selected countries, allow us to define high-level recommendations. These recommendations might be of interest to a diverse array of stakeholders, including government agencies, developers, researchers, and data providers. Government agencies and developers aiming to enhance the usability, collaboration, sustainability, and robustness of their portals can implement these recommendations to optimize their open data initiatives and foster greater citizen engagement through elicitation of requirements for their portal re-engineering. Data providers seeking to improve citizen engagement can engage in direct communication with data consumers, share simplified content, and maintain schema descriptions. The recommendations also address contemporary challenges faced by open data portals, which researchers may find valuable for their investigations. All in all, nineteen recommendations spanning nine sub-dimensions of the developed framework were defined.

**Recommendation 1**: *Provide full English language support ("Multilingualism")*. Portals should provide translations for navigation elements, dataset metadata, articles, manuals, and documentation to enhance English search support and facilitate interaction with an international audience. Exemplary illustrations can be found in the portals of Bahrain, Estonia, Ireland, Malta, Oman, Qatar, Saudi Arabia, and UAE. In addition, the European data portal provides an example of dataset metadata translation.

**Recommendation 2**: *Provide intuitive navigation ("Navigation")*. Navigation elements should be easy to notice, simple, and intuitive. Exemplary illustrations can be found in the portals of Poland, France, Portugal, Saudi

Arabia, Czechia, Germany, and Italy.

**Recommendation 3**: *Ensure consistency ("Navigation", "General performance")*. Portal functionalities should remain consistent, particularly regarding navigation and general performance, but preferably - more broadly. Exemplary illustrations can be found in the portals of Austria, Bahrain, Belgium, Croatia, Finland, France, Estonia, Italy, Latvia, Lithuania, Luxembourg, Netherlands, Poland, Portugal, Slovenia, Spain, and Sweden.

**Recommendation 4**: *Vulgarize content ("Data understandability")*. Simplifying content involves transforming raw data into user-friendly formats and visualizations, making it easier for the general public to understand and utilize the artifacts - data, information, features. This includes creating impact sections, success stories, and displaying use cases, data reuses, applications, and services built on open data. Exemplary illustrations can be found in the portals of the Netherlands, France, Portugal, and Germany.

**Recommendation 5**: *Provide diverse means to visualize data ("Data understandability")*. Offering a range of graphical representations, such as charts, graphs, maps, and interactive displays, helps convey complex information in an easily understandable and engaging format. Exemplary illustrations can be found in the portals of Qatar, Bahrain, Saudi Arabia, Lithuania, Spain, and Poland.

**Recommendation 6**: *Promote High-Value Datasets ("Data understandability")*. Actively showcasing and making easily accessible datasets valuable for research, decision-making, or public interest encourages their utilization and maximizes their impact. This might include additional filtering criteria, promotion on the catalog page, featured lists, or reports highlighting the most valuable datasets. Exemplary illustrations can be found in the portals of Ireland (additional filtering criteria), Lithuania (additional filtering criteria), Slovenia (additional filtering criteria), Poland (additional filtering criteria, promotion), the Netherlands (featured list), and France (reports).

**Recommendation 7**: *Introduce data quality indicators ("Data quality")*. Establishing metrics or criteria to assess the reliability, accuracy, completeness, and consistency of data helps users understand the trustworthiness and usability of datasets. Exemplary illustrations can be found in the portals of France and Portugal.

**Recommendation 8**: *Expose data schema descriptions ("Data quality")*. Providing detailed explanations of the structure and meaning of the various elements within a dataset facilitates comprehension and usage by those unfamiliar with the dataset's underlying structure and terminology. Exemplary illustrations can be found in the portals of Bahrain and Qatar.

**Recommendation 9**: *Expose API/SPARQL endpoints ("Data findability")*. Facilitating seamless data access and querying through API/SPARQL endpoints enables programmatic retrieval and data manipulation for various applications and analyses. Exemplary illustrations can be found in the portals of Ireland, Luxembourg, France, Austria, and Poland.

**Recommendation 10**: *Support complex search prompts ("Data findability")*. Supporting long and complex search prompts empowers users to refine their queries with advanced filters and operators , facilitating precise data retrieval tailored to their needs and enhancing the overall search experience. Exemplary illustrations can be found in the portals of Ireland, Portugal, and Sweden. Given the increasing popularity of Large Language Models use, we can expect more examples to come here.

**Recommendation 11**: *Implement featured topics ("Data findability")*. Highlighting curated collections of datasets enables users to quickly discover and explore most relevant content in various areas. Exemplary illustrations can be found in the portals of France, Italy, the Netherlands, Poland, and Luxembourg.

**Recommendation 12**: *Adopt advanced gamification ("Public engagement")*. Integrating advanced gamification elements, such as competitions, quizzes, rewards, and badges , may enhance engagement, motivation, and participation among users.

**Recommendation 13**: *Notify users about search results' updates ("Public engagement")*. Creating an opportunity to subscribe to dataset or search result updates involves alerting users when new or relevant artifact - data, information - becomes available, ensuring they stay informed. Exemplary illustrations can be found in the Polish portal.

**Recommendation 14**: *Provide comment sections ("Feedback mechanisms & service quality")*. Allowing users to engage with content by sharing their thoughts, opinions, and feedback fosters discussions and stakeholder interaction around the data presented. Exemplary illustrations can be found in the portals of France, Croatia, Lithuania, and Luxembourg.

**Recommendation 15**: *Offer diverse specialized communication channels ("Feedback mechanisms & service quality")*. Offering multiple communication channels enables users to give feedback or receive help through various means such as email, issue-specific forms, chat support, and forums. Exemplary illustrations can be found in the portals of Saudi Arabia, Spain, and Croatia.

**Recommendation 16**: *Introduce usefulness assessment ("Feedback mechanisms & service quality")*. Allowing users to provide feedback on the value of content (e.g., dataset pages, blogs, documentation pages, comments, forum posts) through upvoting or a 5-scale assessment helps improve relevance, usability, and satisfaction. Exemplary illustrations can be found in the portals of Croatia, the Netherlands, Estonia, France, Portugal, and Latvia.

**Recommendation 17**: *Provide dataset release tracking ("Portal sustainability and collaboration")*. Monitoring and documenting the publication of new datasets keeps users informed about the availability of fresh data. Exemplary illustrations can be found in the French portal.

**Recommendation 18**: *Provide user satisfaction surveys ("Portal sustainability and collaboration")*. Offering structured questionnaires or feedback forms gathers valuable insights to improve user experience and address concerns or issues. Exemplary illustrations can be found in the portals of Finland, France, and Saudi Arabia.

**Recommendation 19**: *Emphasize on collaboration ("Portal sustainability and collaboration")*. Initiating and advertising collaboration with other portals, either from the same or different country or region, involves establishing partnerships and sharing resources and data. Exemplary illustrations can be found in the portals of Qatar, with advertisements in the portals of Germany, Austria, and Ireland.

As such, these recommendations highlight contemporary obstacles that open data portals face and suggest strategies to overcome them, benefiting a wide range of stakeholders in the open data ecosystem.

## 6.2   Implications

The study has several theoretical implications. First, we update the criteria and metrics used to assess the user-centricity of open data portals, identifying patterns and trends observed in the literature. These concepts and criteria are used to develop a new integrated framework for evaluating Open Government Data portals, emphasizing user diversity, collaboration, and data understandability. Its application to EU and GCC portals enables comparative analysis of portals from various regions, fulfilling the requirement for an updated viewpoint on benchmarking open data portal performance, which highlights the importance of addressing usability challenges, communication barriers, and strategic value creation in OGD portals to enhance transparency and collaboration among stakeholders.

By applying the framework to 33 OGD portals, we provide insights into the state of understudied OGD portals of the GCC states, providing a new perspective on the state of EU OGD portals, identifying trends in portal design, collaborative initiatives, and areas of improvement for OGD portals in EU and GCC countries, contributing to the development of user-friendly and sustainable portals.

By conducting cluster analyses and deriving best practices and pain points for the portals, the paper underscores the need for continuous evaluation and enhancement of OGD portals to meet evolving user needs and expectations.

The practical implications are the qualitative and quantitative analyses obtained by assessing the portals and conducting the cluster analysis, as well as the defined recommendations that the portal stakeholders can use to develop user-friendly, collaborative, robust, and sustainable portals.

Practical implications, in turn, include recommendations to overcome obstacles that open data portals face, some of them remain unknown, benefiting a wide range of stakeholders in the open data ecosystem. These include enhancing data descriptions, promoting high-value datasets, incorporating visualization tools, and providing multilingual support. We also suggest implementing features such as query recommendation systems, automatic dataset descriptions, gamification elements, and storytelling to attract a wider audience and enhance content understandability in OGD portals.

By identifying best practices in the OGD portal design, as identified during the qualitative analysis of the above, we also advocate for the adoption of best practices identified through portal assessments.

Finally, although this study presents a comparative analysis of EU and GCC portals, the framework developed herein holds significant value for OGD portal owners and managers. It serves as a practical tool, acting as a checklist for internal assessments of their portals. By using this framework, owners and managers can systematically evaluate their portals' performance and pinpoint areas in need of design or redesign. For the later, they can find best practices within all dimensions that can navigate them in planning respective activities.

## 6.3   Limitations

This study is not without its limitations. One limitation lies in the framework's structure, particularly in combining sub-dimensions into dimensions. This process can sometimes lead to overlapping or conflicting criteria. The difficulty arises from integrating dimensions based on existing frameworks, where similar purposes may result in identical sub-dimensions allocated to different dimensions. To mitigate this limitation, we based our framework on the transparency-by-design framework (Lnenicka & Nikiforova, 2021). Despite our efforts, the potential combinations of sub-dimensions into dimensions remain practically limitless.

Determining the optimal number of sub-dimensions to balance specificity and generality within the framework poses another challenge. This may introduce subjectivity into evaluations. Our aim was to strike a balance between specificity and generality, making the framework applicable for future use without becoming overly general. For example, suggestions arose during the study and expert consultations to include checks for language support in documentation for the "Multilingualism" dimension or specific data visualization types in the "Data understandability" dimension. However, we chose to exclude these to prevent overly granular evaluation criteria.

To maintain future applicability, the framework primarily employs Boolean assessment. This necessitated structuring criteria to elicit yes/no responses, limiting the expression of a third option. Qualitative analysis was used to address this limitation. Additionally, the selection of the weighting system was guided by the need to strike a balance between simplicity and complexity.

The study's sampling of datasets for assessing sub-dimensions is exploratory and may not fully represent all portal features and functionalities. To mitigate this, we devised a sampling strategy to select fourteen datasets inclusively (least and most popular/recent etc.). This approach aimed to provide an equitable opportunity for data exploration while adhering to time and resource constraints.

Furthermore, variations in feature implementation across different portals restricted the evaluation to certain scenarios. This limitation may have restricted the assessment's coverage of all possible portal functionalities and user interactions.

## 7   Conclusion

This study develops an integrated usability framework for evaluating open data portals that focuses on: (1) the portal's ability to adapt to a diverse user base; (2) promoting user collaboration and participation; and (3) enabling users to understand and explore the data. This study introduces an integrated usability framework designed to evaluate open data portals, focusing on their adaptability to diverse users, promotion of collaboration, and facilitation of data understanding and exploration. The framework, comprising 72 dimensions, was developed and applied to assess 33 EU and GCC OGD portals. As a result of its application, (a) each individual portal was evaluated, (b) statistics about the portal performances were compiled, (c) EU and GCC portals were ranked, (d) best practices and pain points were identified, (e) trends in portal design and collaborative initiatives between portals at the intra- and interregional levels were identified. These findings informed the derivation of 19 high-level recommendations aimed at addressing common challenges in the open data ecosystem, benefiting stakeholders across sectors.

Portal assessments involved computing sub-dimension, dimension, and total scores using a weighing system, culminating in average scores for EU and GCC portals and the identification of top and low performers. Rankings were established based on individual dimension results and total scores, alongside qualitative analysis pinpointing best practices and pain points through portal-specific examples.

The notable performance of top European portals (based on EU Open Data Maturity Reports) within this framework suggests alignment with existing benchmarks, albeit with unique considerations. Notably, Saudi Arabian, Qatari, and Bahraini portals demonstrated competence, even setting trends in certain sub-dimensions. The analysis underscores the importance of enhancing multilingual support, user communication channels, and dataset usability to foster engagement and repeated portal use.

A growing trend towards exposing data quality indicators and involving users in portal ecosystems was observed, emphasizing the need for transparent feature highlighting and the incorporation of advanced functionalities like assistants, AI-augmented recommender systems, advanced search, incl. NLP or LLM capabilities for advanced search or examining datasets, as well as gamification elements. Effective feedback mechanisms can enhance user participation and dataset quality, with the framework serving as a baseline requirement for OGD portals.

This framework should be revisited once there are examples of how the above technologies can be advantageous for open data portals. Currently, the framework can be seen as the "*minimum set of requirements*" that the OGD portal must comply with.

Comparative analysis of portals from different regions reveals insights into diverse implementation approaches, while examples of cross-regional collaboration underscore its potential to enrich the open data ecosystem. Future research avenues may explore expanding framework dimensions, evaluating language barrier impacts, and periodically reassessing portals to monitor evolving trends and advancements. This holistic approach contributes to a deeper understanding of open data ecosystem dynamics at national and interregional levels.

# 8    Declaration of Generative AI and AI-assisted technologies in the writing process

During the preparation of this work the authors used ChatGPT-3.5 in order to improve the wording and streamline selected text fragments. After using this tool/service, the authors reviewed and edited the content as needed and takes full responsibility for the content of the publication.

# Appendix

# I. Portals web addresses

| Country | Web address |
| --- | --- |
| Austria | www.data.gv.at |
| Bahrain | www.data.gov.bh/pages/homepage/ |
| Belgium | data.gov.be/en |
| Bulgaria | data.egov.bg/ |
| Croatia | data.gov.hr/en |
| Cyprus | www.data.gov.cy/?language=en |
| Czechia | data.gov.cz/english/ |
| Denmark | www.opendata.dk/ |
| Estonia | avaandmed.eesti.ee/ |
| Finland | www.avoindata.fi/en |
| France | data.gouv.fr |
| Germany | www.govdata.de/ |
| Greece | www.data.gov.gr/ |
| Hungary | kozadatportal.hu/ |
| Ireland | data.gov.ie |
| Italy | www.dati.gov.it/ |
| Kuwait | e.gov.kw/sites/kgoenglish/Pages/OtherTopics/OpenData.aspx |
| Latvia | data.gov.lv/eng |
| Lithuania | data.gov.lt/ |
| Luxembourg | data.public.lu/en/ |
| Malta | data.gov.mt/ |
| Netherlands | data.overheid.nl/en |
| Oman | data.gov.om/ |
| Poland | dane.gov.pl/en dados.gov.pt/en/ |
| Portugal | |
| Qatar | www.data.gov.qa/pages/home/ |
| Romania | data.gov.ro/ |
| Saudi Arabia | od.data.gov.sa/en |
| Slovakia | data.gov.sk/en |
| Slovenia | podatki.gov.si/# |
| Spain | datos.gob.es/ |
| Sweden | www.dataportal.se/en |
| United Arab Emirates | bayanat.ae/ |

Table 8. Portals web addresses